\documentclass[]{pasj02} 
\usepackage[switch,mathlines]{lineno} 
\usepackage{natbib} 
\usepackage{color}

\DeclareAbbreviation\jcap{JCAP}

\jyear{2024}
\Received{}
\Accepted{}

\begin{document} 

\title{Catalogs of optically-selected clusters and photometric
  luminous red galaxies from the Hyper Suprime-Cam Subaru Strategic
  Program final year dataset}

\author{
 Masamune \textsc{Oguri},\altaffilmark{1,2}\altemailmark\orcid{0000-0003-3484-399X} \email{masamune.oguri@chiba-u.jp} 
Yen-Ting \textsc{Lin},\altaffilmark{3}\orcid{0000-0001-7146-4687}
Nobuhiro \textsc{Okabe},\altaffilmark{4,5,6}\orcid{0000-0003-2898-0728}
Naomi \textsc{Ota},\altaffilmark{7}\orcid{0000-0002-2784-3652}
I-Non \textsc{Chiu},\altaffilmark{8}\orcid{0000-0002-5819-6566}
Momoka \textsc{Fujikawa},\altaffilmark{2}
Hung-Yu \textsc{Jian},\altaffilmark{3}\orcid{0000-0002-6121-8420}
Tadayuki \textsc{Kodama},\altaffilmark{9}\orcid{0000-0002-2993-1576}
Lihwai \textsc{Lin},\altaffilmark{3}\orcid{0000-0001-7218-7407}
Atsushi J. \textsc{Nishizawa},\altaffilmark{10,11}\orcid{0000-0002-6109-2397}
Rhythm \textsc{Shimakawa},\altaffilmark{12}\orcid{0000-0003-4442-2750}
Yoshiki \textsc{Toba},\altaffilmark{13,3,14}\orcid{0000-0002-3531-7863}
}
\altaffiltext{1}{Center for Frontier Science, Chiba University, 1-33 Yayoi-cho, Inage-ku, Chiba 263-8522, Japan}
\altaffiltext{2}{Department of Physics, Graduate School of Science, Chiba University, 1-33 Yayoi-Cho, Inage-Ku, Chiba 263-8522, Japan}
\altaffiltext{3}{Institute of Astronomy \& Astrophysics, Academia Sinica, No.1, Sec. 4 Roosevelt Road, Taipei 106216, Taiwan}
\altaffiltext{4}{Physics Program, Graduate School of Advanced Science and Engineering, Hiroshima University, 1-3-1 Kagamiyama, Higashi-Hiroshima, Hiroshima 739-8526, Japan}
\altaffiltext{5}{Hiroshima Astrophysical Science Center, Hiroshima University, 1-3-1 Kagamiyama, Higashi-Hiroshima, Hiroshima 739-8526, Japan}
\altaffiltext{6}{Core Research for Energetic Universe, Hiroshima University, 1-3-1, Kagamiyama, Higashi-Hiroshima, Hiroshima 739-8526, Japan}
\altaffiltext{7}{Department of Physics, Nara Women's University, Kitauoyanishi-machi, Nara, Nara 630-8506, Japan}
\altaffiltext{8}{Department of Physics, National Cheng Kung University, No.1, University Road, Tainan City 70101, Taiwan}
\altaffiltext{9}{Astronomical Institute, Tohoku University, 6-3 Aramaki, Aoba-ku, Sendai 980-8578, Japan}
\altaffiltext{10}{DX Center, Gifu Shotoku Gakuen University, Takakuwanishi, Yanaizu, Gifu, 501-6194, Japan}
\altaffiltext{11}{Kobayashi Maskawa Institute, Nagoya University, Furocho, Chikusa, Nagoya, Aichi, 464-8602, Japan}
\altaffiltext{12}{Waseda Institute for Advanced Study (WIAS), Waseda University, 1-21-1, Nishi-Waseda, Shinjuku, Tokyo 169-0051, Japan}
\altaffiltext{13}{Department of Physical Sciences, Ritsumeikan University, 1-1-1 Noji-higashi, Kusatsu, Shiga 525-8577, Japan}
\altaffiltext{14}{Research Center for Space and Cosmic Evolution, Ehime University, 2-5 Bunkyo-cho, Matsuyama, Ehime 790-8577, Japan}



\maketitle

\begin{abstract}
We construct samples of optically-selected clusters and photometric
luminous red galaxies (LRGs) from the Hyper Suprime-Cam Subaru
Strategic Program final year dataset covering $\sim 1200$~deg$^2$.
The cluster catalogs extend out to the redshift of $1.38$ and contain
more than 10000 clusters with richness larger than $15$, where
the richness is defined to be a membership probability weighted
number of galaxies above the stellar masses of approximately
$10^{10.2}M_\odot$. The total number of probable red cluster member
galaxies in these clusters are more than $6\times 10^5$. Photometric
redshifts of the clusters are shown to be precise with the scatter
better than $\sim 0.01$ for a wide redshift range.  We detect stacked
weak lensing signals of clusters out to the  redshift of $1$, and use
them to update constraints on the mass-richness relation. Our catalog
of about 6 million photometric LRGs extend out to the redshift of
$1.25$, and have the scatter of the photometric redshift better than
$\sim 0.02$ for the redshift range between $0.4$ and $1.0$. 
\end{abstract}


\section{Introduction}

Clusters of galaxies are the most massive gravitationally bound
objects in the Universe. Their abundance and internal structure are
determined mainly by the gravitational dynamics, which suggests that
clusters serve as a useful testbed for theories of gravity and
cosmology. For instance, the abundance of clusters has been used to
place tight constraints on cosmological parameters
\citep[e.g.,][for recent examples]{ghirardini2024,sunayama2024,chiu2024,des2025}.
In addition, clusters of galaxies serve as a useful laboratory to
study the environment effect of galaxy formation and evolution as well
as diffuse plasma at extreme energy scales
\citep[e.g.,][for a review]{kravtsov12}. 

Because clusters of galaxies are dominated by dark matter that is not
directly visible, they have been identified with several different
methods. The most traditional approach is to select clusters of
galaxies from spatial concentrations of galaxies. Since clusters of galaxies
usually contain a huge amount of hot gas, they can be identified by
X-ray observations as well as the Sunyaev-Zel'dovich effect in the
centimeter to millimeter wavelength range. See e.g.,
\citet{weinberg13} for a comparison of selection functions of cluster
samples constructed in different wavelengths. Recently, it has become
possible to construct sample of clusters directly using weak
gravitational lensing \citep[see][for a review]{oguri25}. 

Clusters of galaxies are efficiently identified from concentrations of
red galaxies
\citep[e.g.,][]{galdders00,koester07,hao10,szabo11,wen12,rykoff14,oguri14,bellagamba18,gonzalez19,grishin23}.
An advantage of optically-selected cluster samples is that accurate
photometric redshifts are available from multi-wavelength imaging
data. Thanks to the progress of wide-field galaxy surveys in optical and
near-infrared bands, large samples of optically-selected clusters have
been constructed out to high redshifts. For instance, from the Dark
Energy Survey (DES) Year 3 data, \citet{des2025} construct a volume-limited
sample of optically selected clusters out to $z=0.65$. From the Kilo
Degree Survey (KiDS) Data Release 3, \citet{maturi19} construct a sample of 
optically-selected clusters at $0.1<z<0.8$
\citep[see also][]{maturi25}. \citet{wen24} construct a
large sample of more than a million clusters out to $z\sim 1$ based on
the DESI Legacy Imaging Surveys
\citep[see also][]{yang21,zou21,yantovski24}. The recent analysis of
the Euclid Quick Data Release 1 by \citet{euclid2025} demonstrates
that clusters out to $z\sim 1.5$ can easily be identified in the
Euclid imaging data. 

The Hyper Suprime-Cam Subaru Strategic Program \citep[HSC-SSP;][]{aihara18a}
provides the deepest imaging data among the so-called Stage-III galaxy
surveys, which enables the identification of clusters at $z\sim 1$ and
beyond. Adopting the CAMIRA cluster finding algorithm \citep{oguri14}
that makes use of the stellar population synthesis model of
\citet{bruzual03} to model the red-sequence, \citet{oguri18a}
construct an optically-selected cluster catalog out to the redshift of
$1.1$ from the first-year (S16A) HSC-SSP dataset covering
$\sim 232$~deg$^2$. The mass-richness relation of CAMIRA clusters are
calibrated by weak gravitational lensing \citep{murata19,chiu20a} and
clustering \citep{chiu20b}. Follow-up observations with X-ray \citep{ota20} and
the Sunyaev-Zel'dovich effect \citep{okabe21,kitayama23} confirm that
the hot gas is indeed associated with rich CAMIRA clusters. 
As the HSC-SSP survey progresses, the CAMIRA cluster
catalog is continuously updated
\citep[e.g.,][]{willis21,oguri21,klein22,hashiguchi23,toba24,chen24}.
\citet{wen21} construct an independent cluster catalog from the HSC-SSP out to
$z\sim 2$ in combination with WISE near-infrared data \citep{wright10}. 

In this paper, we present CAMIRA cluster catalogs from the HSC-SSP
final year (S23B) dataset. The catalogs contain more than 10,000
clusters in the wide photometric redshift range of $0.1\leq z\leq 1.38$,
$\sim 3000$ of which are located at $z\geq 0.8$. The CAMIRA cluster
catalogs extend to higher redshifts than those in DES, KiDS, and DESI
Legacy Imaging Surveys, and serve as a stepping stone to Euclid
cluster catalogs that will also contain a large number of clusters at
$z\gtrsim 1$. We study the properties of our cluster catalogs,
including the mis-centering and the mass-richness relation. In
addition to the cluster catalogs, we construct catalogs of photometric
luminous red galaxies (LRGs), which are byproducts of the CAMIRA
algorithm to fit galaxy colors and redshifts. The photometric LRG
catalog from the first-year (S16A) HSC-SSP dataset is presented in
\citet{oguri18b} and its properties are studied in e.g.,
\citet{ishikawa21} and \citet{rau23}. The updated photometric LRG
catalogs are constructed also from the HSC-SSP final year (S23B)
dataset and contain $\sim 6$ million photometric LRGs in the
photometric redshift range of $0.05\leq z\leq 1.25$.
Compared with the S16A catalogs published in \citet{oguri18a} and
 \citet{oguri18b}, the final year cluster and photometric LRG catalogs
 presented in this paper contain roughly a factor of 5 more objects,
which enable much more precise studies of the galaxy evolution and
cosmology.

This paper is organized as follows. Sec.~\ref{sec:method} describes
the method and data that are used to construct the catalogs.
Sec.~\ref{sec:cluster} describes optically-selected cluster
catalogs and their basic characteristics. Sec.~\ref{sec:lrg}
describes photometric LRG catalogs. Finally we summarize our results
in Sec.~\ref{sec:summary}. Catalogs will be made available at
{\tt https://github.com/oguri/cluster\_catalogs}.

\section{Data and method}\label{sec:method}

\subsection{HSC-SSP final year dataset}\label{sec:hsc-ssp}

The HSC-SSP is a wide-field optical imaging survey with the Hyper
Suprime-Cam \citep[HSC;][]{miyazaki18a,komiyama18,furusawa18} mounted on
the 8.2-meter Subaru Telescope. The HSC-SSP consists of three layers,
Wide, Deep, and Ultradeep, and the Wide layer takes images in five
($grizy$) broadbands \citep{kawanomoto18} over more than 1,000~deg$^2$
with a limiting magnitude of $i_{\mathrm{lim}}\sim 26$ at
$5\sigma$ for point sources with a $2''$ diameter aperture.
The HSC-SSP observations are conduced in 2014--2021 and the 
data are reduced with hscPipe \citep{bosch18}. The HSC-SSP data are
made publicly available through a series of data releases
\citep{aihara18b,aihara19,aihara22}.

The HSC-SSP final year dataset is internally referred to as S23B and
will correspond to Public Data Release 4. From the HSC-SSP database,
we select galaxies with 
{\tt isprimary}~$=1$ and the star-galaxy separation parameter
{\tt i\_extendedness\_value}~$= 1$ and that are observed in all the five
broadbands with the numbers of visits {\tt inputcount\_value}~$\geq 2$ for
$gr$-bands and {\tt inputcount\_value}~$\geq 3$ for $izy$-bands. We remove
any galaxies with any of the following flags in any of the five
broadbands; {\tt pixelflags\_edge}, {\tt
  pixelflags\_interpolatedcenter}, and {\tt pixelflags\_crcenter}. We
use galaxies with $z$-band {\tt cmodel} magnitudes after the Galactic
dust extinction corrections \citep{schlegel98} brighter than $z$-band
magnitudes of $24$.
Additionally we remove galaxies with $z$-band {\tt cmodel} magnitude
error larger than $0.1$. As in \citet{oguri18a}, we use {\tt cmodel}
magnitudes for magnitudes of galaxies, while the point spread function
(PSF)-matched aperture photometry with the target PSF size of
$1\farcs3$ and the aperture of $1\farcs5$ in diameter to avoid the
deblending issue \citep{aihara18b}. See Appendix~\ref{app:sql_query}
for an example of the SQL query for constructing our parent galaxy
sample. 

We generate galaxy catalogs both with and without the bright star mask
\citep{coupon18,aihara22}. When applying the bright star mask, we
remove galaxies with any of the following flags in any of the five
broadbands; {\tt mask\_brightstar\_halo}, {\tt
  mask\_brightstar\_ghost}, and {\tt mask\_brightstar\_blooming}.
We extract galaxy catalogs from the Wide layer as well as the Deep and
Ultradeep (hereafter Deep+UD) layers, where we use the same $z$-band
limiting magnitude of $24$ for both Wide and Deep+UD.

\subsection{Spectroscopic catalogs}\label{sec:spec_z}

The CAMIRA cluster finding algorithm uses spectroscopic galaxies for
the purpose of calibrating red-sequence galaxy colors
\citep{oguri14,oguri18a}. We use spectroscopic galaxies compiled in the
HSC-SSP database, which include spectroscopic galaxies from
zCOSMOS DR3 \citep{lilly07}, 3D-HST v4.1.5 \citep{skelton14,momcheva16}, SDSS DR15
\citep{aguado19}, GAMA DR3 \citep{baldry18}, UDSz DR1\citep{bradshaw13,malure13}, 
VANDELS DR4 \citep{garilli21}, C3R2 DR2 \citep{masters19}, VVDS final
data release \citep{lefevre13}, DEIMOS 10k \citep{hasinger18},
FMOS-COSMOS \citep{silverman15}, LEGA-C DR2, \citep{straatman18},
PAUS+COSMOS v0.4 \citep{alarcon21}, PRIMUS DR1 \citep{coil11,cool13},
VIPERS DR2 \citep{scodeggio18}, WiggleZ DR1 \citep{drinkwater10},
DEEP2 \citep{newman13}, and DEEP3 \citep{cooper12}. In addition, we
include spectroscopic galaxies from GOGREEN and GCLASS First Data
Release \citep{balogh21}, follow-up spectroscopic observations of
the CL1604 supercluster \citep{hayashi19}, and our own follow-up
spectroscopic observations of high-redshift clusters using the
Magellan Telescope. Furthermore, we include a large sample of
spectroscopic galaxies from DESI DR1 \citep{2025arXiv250314745D} with
{\tt ZWARN}~$=0$ and {\tt ZCAT\_PRIMARY}~$=1$.

We cross match these spectroscopic galaxies with photometric galaxies
in the HSC-SSP with the bright star mask (see Sec.~\ref{sec:hsc-ssp}),
using a matching separation tolerance of $1''$. We remove any
spectroscopic galaxies with measurement error larger than
$0.01(1+z_{\mathrm{spec}})$. The numbers of spectroscopic galaxies that
are cross-matched are 2925976 for Wide and 172930 for Deep+UD. We then
select spectroscopic galaxies at $0.01\leq z\leq 1.4$, and apply color cuts
summarized in Appendix~\ref{app:color_cuts} for an approximate
selection of red-sequence galaxies, which results in 915571 galaxies
combining those in the Wide and Deep+UD layers. These spectroscopic
galaxies are used for the CAMIRA cluster finding algorithm to refine
a stellar population synthesis model of red-sequence galaxies.

\subsection{CAMIRA algorithm}\label{sec:algorithm}

We mostly follow the methodology to run the CAMIRA algorithm presented
in \citet{oguri14} and \citet{oguri18a}. The CAMIRA algorithm fits all
photometric galaxies with a stellar population synthesis model of
\citet{bruzual03} with the calibration of galaxy colors using the
spectroscopic galaxy sample constructed in Sec.~\ref{sec:spec_z}.
We assume a single instantaneous burst at a formation redshift of
$z=3$, and the stellar mass dependence of the metallicity is included
to reproduce a tilt of the color-magnitude relation of red-sequence
galaxies. The performance of the algorithm is expected not to
  depend on the specific choice of the assumed star formation history
  because of the calibration of galaxy colors with spectroscopic
  galaxies.  The likelihood of the fitting for each galaxy as a function
of redshift is used to construct the three-dimensional richness map.
In the CAMIRA algorithm, the richness is defined by the number of
  red-sequence galaxies with stellar masses $M_*\gtrsim
  10^{10.2}M_\odot$ and within a circular aperture with a physical
  radius of $R\lesssim 1 h^{-1}\mathrm{Mpc}$. In calculating the
  richness, CAMIRA uses a smoothly truncated stellar mass and spatial
  filters and the number of galaxies is estimated by the sum of
  likelihoods of being the red-sequence galaxies computed for
  individual galaxies with the stellar population synthesis fitting
  \citep[see][for more details]{oguri14}. We note
that the stellar initial mass function of \citet{salpeter55} is
assumed in computing the stellar mass. We use a compensated spatial
filter in constructing the richness map such that the background
levels are automatically subtracted. For each peak in the richness
map, a bright galaxy near the peak is selected as the
brightest cluster galaxy (BCG). The richness $\hat{N}_{\mathrm{mem}}$
and the photometric redshift $z_{\mathrm{cl}}$ are re-computed with
the identified BCG as the center. Interested readers are referred to
\citet{oguri14} and \citet{oguri18a} for more details on the procedure
and specific choices of parameters used in the CAMIRA algorithm.

As a byproduct, photometric LRG catalogs are also constructed taking
advantage of the CAMIRA algorithm. Specifically, the stellar 
population synthesis model of \citet{bruzual03} with the calibration
of galaxy colors by the spectroscopic galaxies is used to fit all the
photometric galaxies used for finding clusters (see
Sec.~\ref{sec:hsc-ssp} for the selection) leaving the redshift and
stellar mass as free parameters. We then select galaxies with
best-fitting chi-square of $\chi^2<10$ as photometric LRGs. Since the
stellar population synthesis model used for the CAMIRA algorithm
include red galaxies only, we can select a sample of red galaxies with
little star formation by this procedure.   

\begin{table}
  \tbl{HSC-SSP final year CAMIRA cluster catalogs\footnotemark[$*$] }{%
  \begin{tabular}{ccc}
      \hline
      Catalog & Area [deg$^2$] & \# of clusters\\ 
      \hline
      Wide w/ star mask    & 1220 & 12162 \\
      Wide w/o star mask   & 1289 & 12939 \\
      Deep+UD w/ star mask & 29 & 238   \\
      Deep+UD w/o star mask& 33 & 268   \\
      \hline
    \end{tabular}}\label{tab:clusters}
\begin{tabnote}
\footnotemark[$*$] The redshift range is $0.1\leq z_{\mathrm{cl}}\leq 1.38$ and the richness range is $\hat{N}_{\mathrm{mem}}\geq 15$ for all the cluster catalogs 
\end{tabnote}
\end{table}

\begin{figure}
 \begin{center}
   \includegraphics[width=8.6cm]{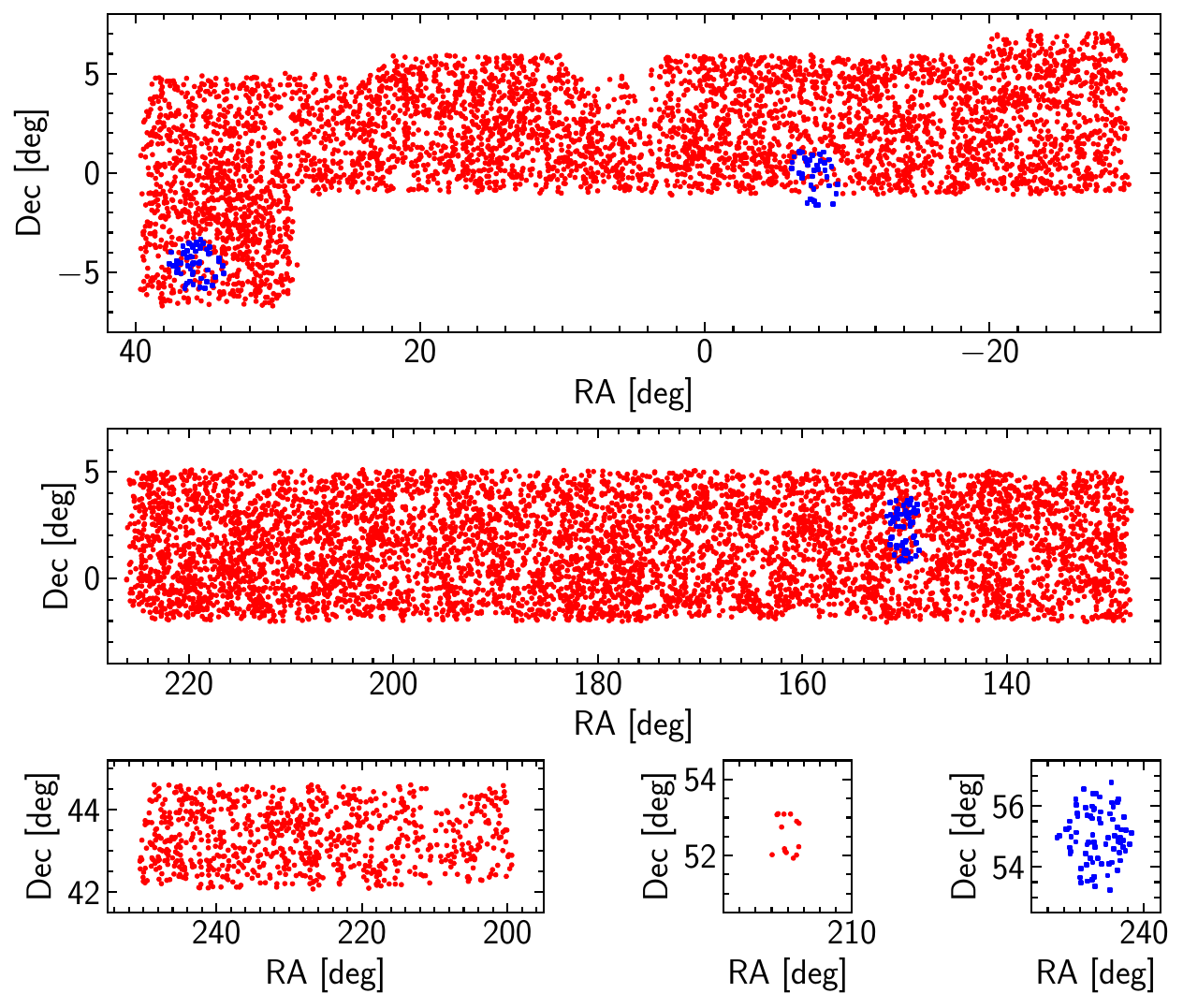} 
 \end{center}
\caption{Distribution of CAMIRA clusters on the sky. Filled red
  circles show positions of clusters in the Wide layer, while filled
  blue squares show positions of clusters in the Deep+UD layer. The
  bright star mask is applied.
  {Alt text: The plot shows how CAMIRA clusters are distributed on the sky.}
}\label{fig:footprint}
\end{figure}

\section{Cluster catalogs}\label{sec:cluster}

\subsection{Basic characteristics}

Table~\ref{tab:clusters} shows the summary of numbers of clusters in
each catalog. The total numbers of cluster member galaxies with the
membership weight factor $w_{\mathrm{mem}}\geq 10^{-3}$
\citep[see][for the definition]{oguri14} are more than $6\times 10^5$
for the Wide and more than $10^4$ for Deep+UD. We note that no
  cut on the membership weight factor is applied when deriving the
  richness of individual clusters.
Figure~\ref{fig:footprint} shows distributions of clusters
on the sky from the Wide and Deep+UD catalogs when the bright star
mask is applied. Unless otherwise specified, in what follows we always use the
Wide cluster catalog with the bright star mask for the analysis. 

\begin{figure}
 \begin{center}
   \includegraphics[width=8.6cm]{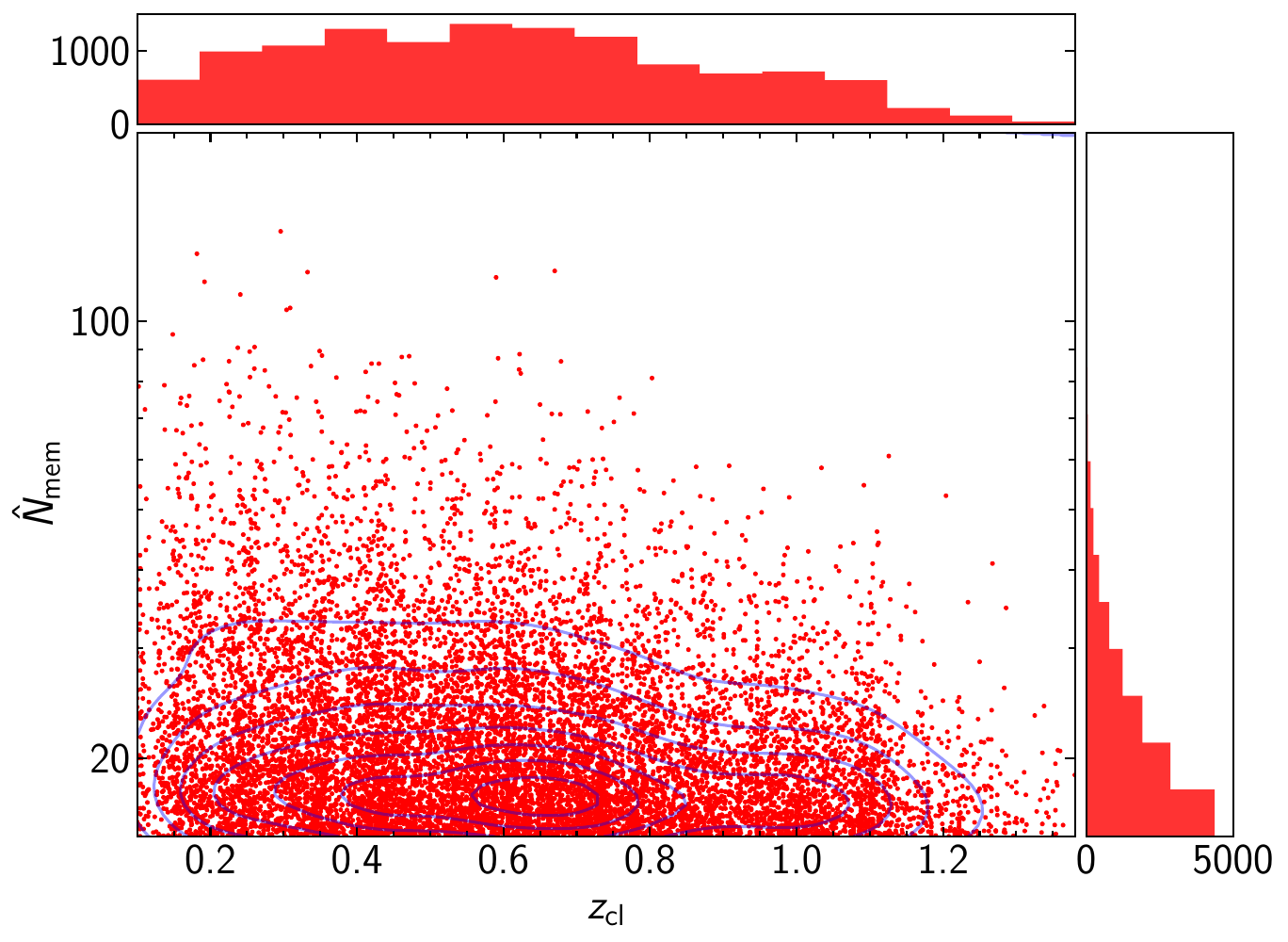} 
 \end{center}
\caption{Distribution of CAMIRA clusters in the redshift-richness
  plane. Contours indicates the density of clusters.
  Top and right panels show the histograms of the cluster
  redshift $z_{\mathrm{cl}}$ and the richness
  $\hat{N}_{\mathrm{mem}}$, respectively.
  {Alt text: In the main plot, the x axis shows the redshift from 0.1
    to 1.38, and the y axis shows the richness from 15 to 200. Points
    show how CAMIRA clusters are distributed in this plane.}
}\label{fig:camira_nz} 
\end{figure}

\begin{figure}
 \begin{center}
   \includegraphics[width=8.6cm]{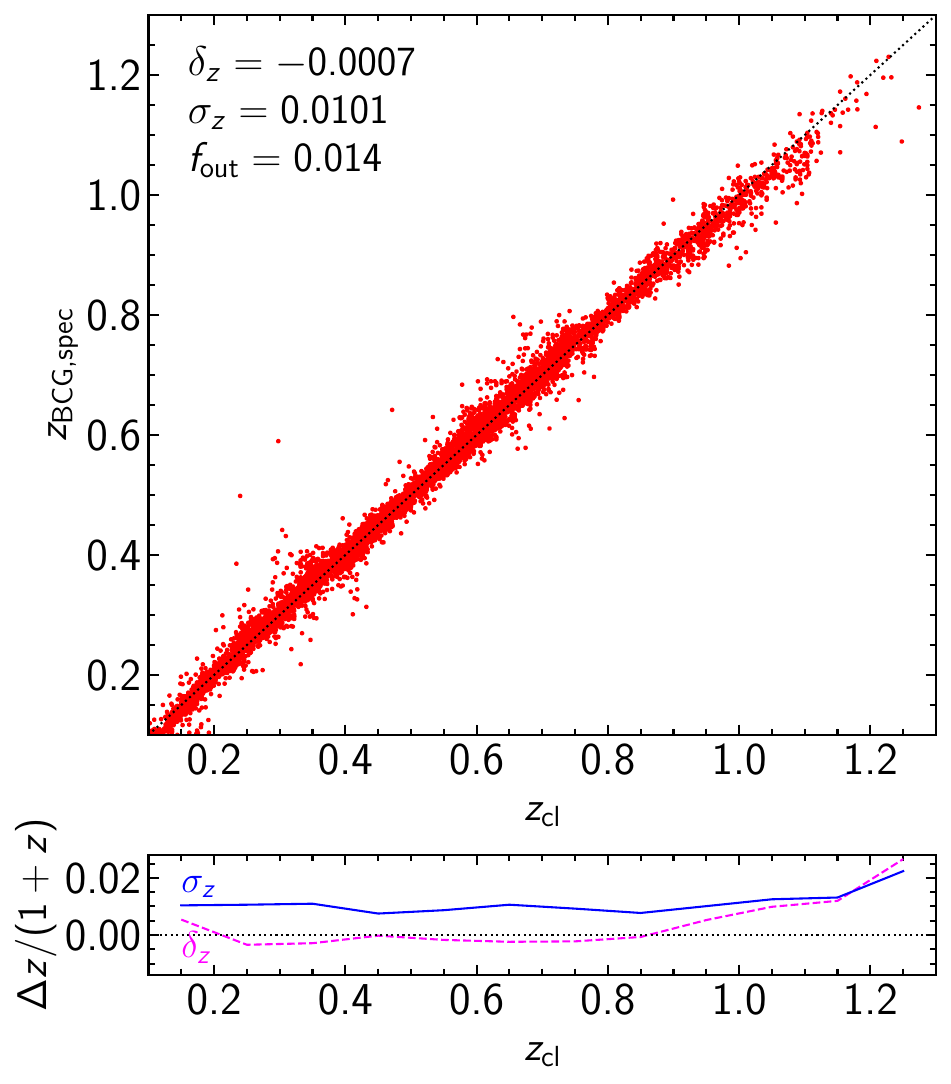} 
 \end{center}
\caption{{\it Upper:} The comparison between cluster photometric redshifts
  $z_{\mathrm{cl}}$ and spectroscopic redshifts of BCGs
  $z_{\mathrm{BCG,spec}}$. Values of the bias $\delta_z$, the scatter
  $\sigma_z$, and the outlier fraction of the redshift residual
  $(z_{\mathrm{cl}}-z_{\mathrm{BCG,spec}})/(1+z_{\mathrm{BCG,spec}})$
  that are derived with $4\sigma$ clipping are also shown. 
  {\it Lower:} The  bias $\delta_z$, the scatter $\sigma_z$ of the
  redshift residual as a function of redshift.
  {Alt text: the x and y axes show the redshift from 0.1  to 1.38.}
}\label{fig:camira_zcomp} 
\end{figure}

The distribution of CAMIRA clusters in the redshift-richness plane is
shown in Figure~\ref{fig:camira_nz}.  It is found that
many clusters of galaxies are identified out to $z\sim 1$. The number
of the richest clusters declines with increasing redshift, reflecting
a rapid evolution of the number density of massive dark matter
halos. We note that the selection of cluster member galaxies used for
computing the richness, which correspond to the stellar mass limit of
$M_*\gtrsim 10^{10.2}M_\odot$, is almost complete out to $z\sim 1.1$
thanks to the exquisite depth of the HSC-SSP
\citep{nishizawa18}. However, at the highest redshift $z\gtrsim 1.1$,
the member galaxy selection becomes slightly
incomplete and as a result the richness is slightly underestimated at
$z\gtrsim 1.1$. Figure~\ref{fig:camira_zcomp} checks the performance
of cluster photometric redshifts by comparing them with spectroscopic
of BCGs. We find that the photometric redshifts are accurate out to
$z_{\mathrm{cl}}\sim 1$. The scatter $\sigma_z$ of the redshift residual
$(z_{\mathrm{cl}}-z_{\mathrm{BCG,spec}})/(1+z_{\mathrm{BCG,spec}})$ is
$\sigma_z\sim 0.01$ for a wide range of the cluster redshift.
At $z_{\mathrm{cl}}\gtrsim 1$ we see a sign of the degradation of the
photometric redshift precision and accuracy, probably because of the
small number of spectroscopic redshifts of red-sequence galaxies for
the calibration as well as the lack of near-infrared bands for accurate
characterizations of the spectral energy distribution of red-sequence
galaxies. 

\begin{figure*}
 \begin{center}
   \includegraphics[width=5.8cm]{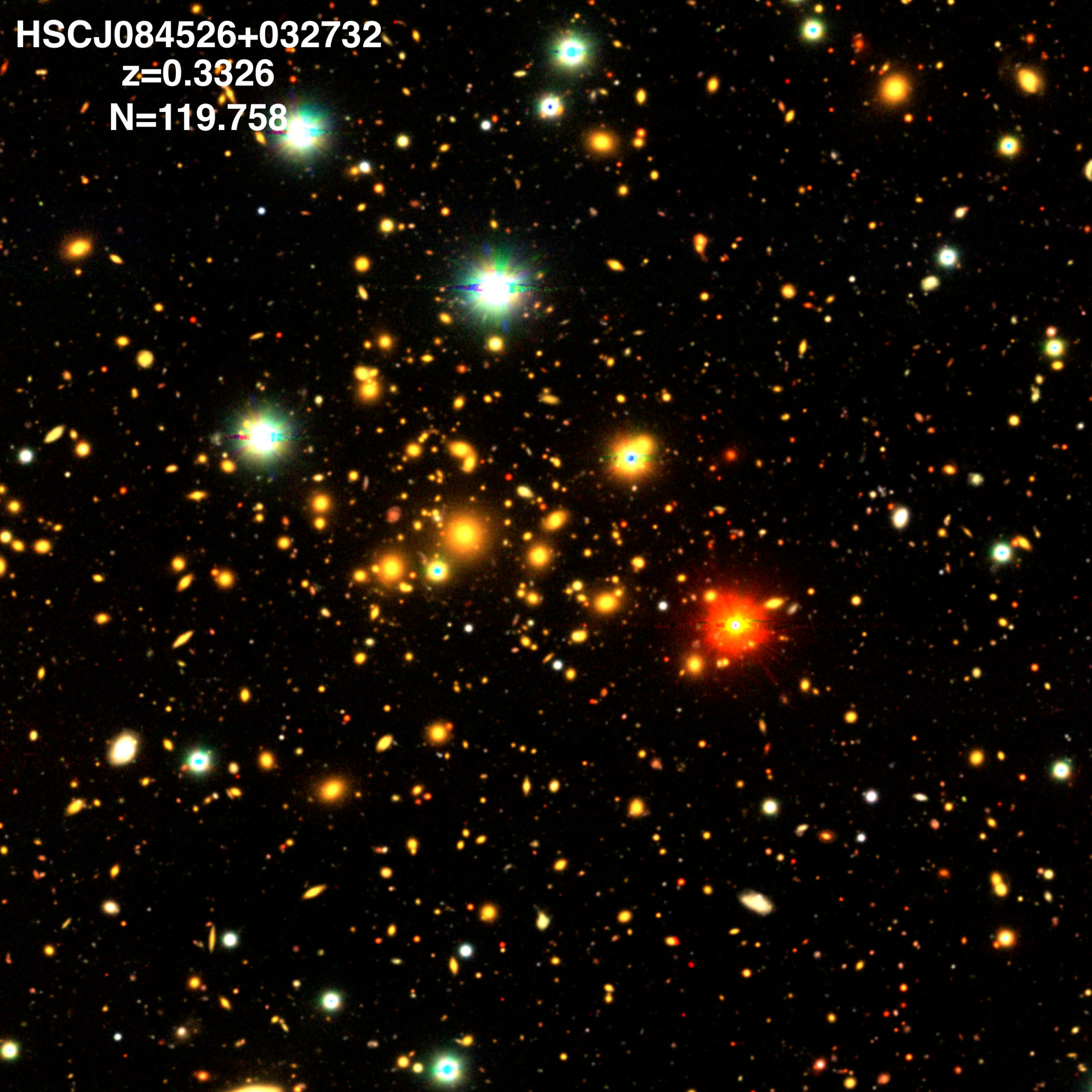} 
   \includegraphics[width=5.8cm]{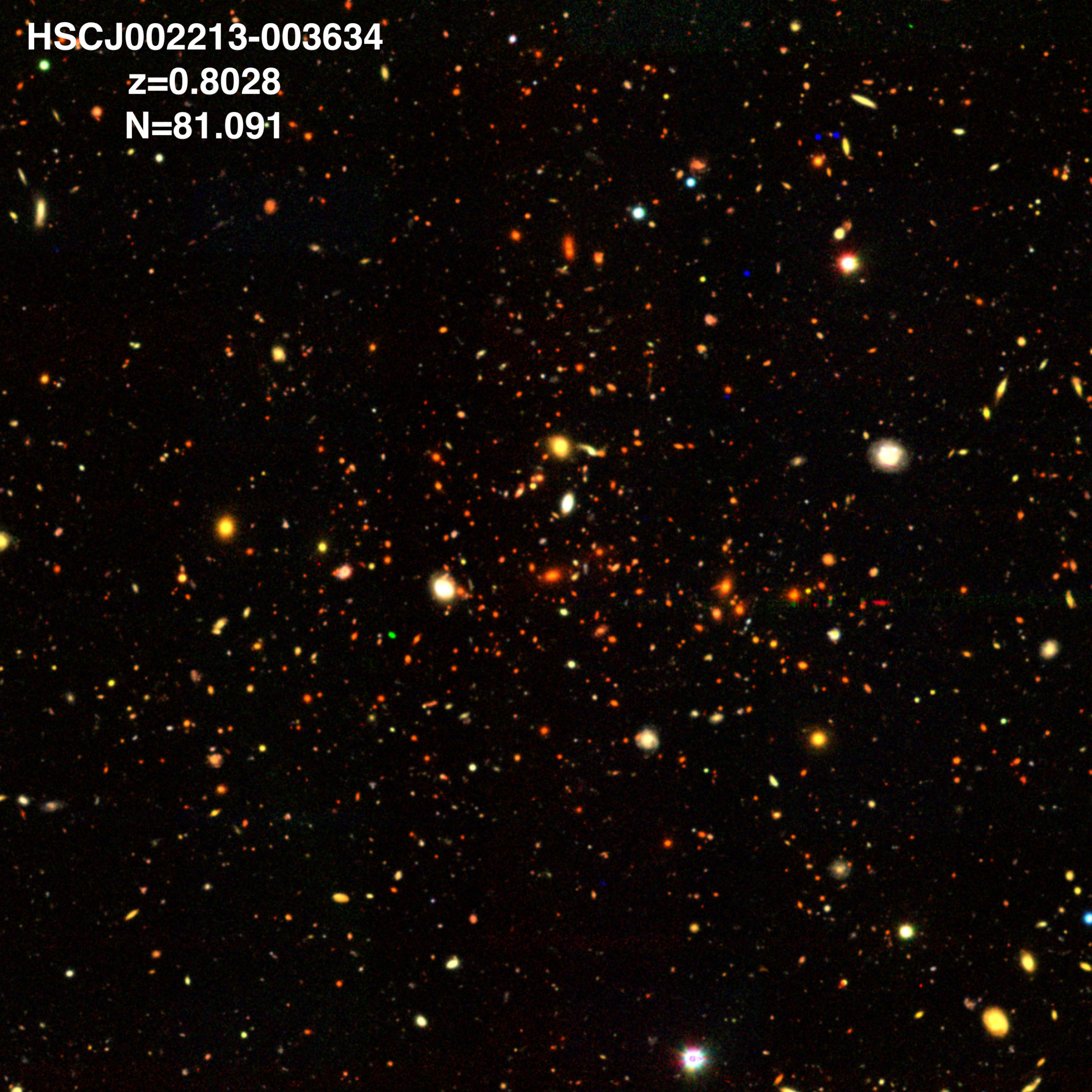} 
   \includegraphics[width=5.8cm]{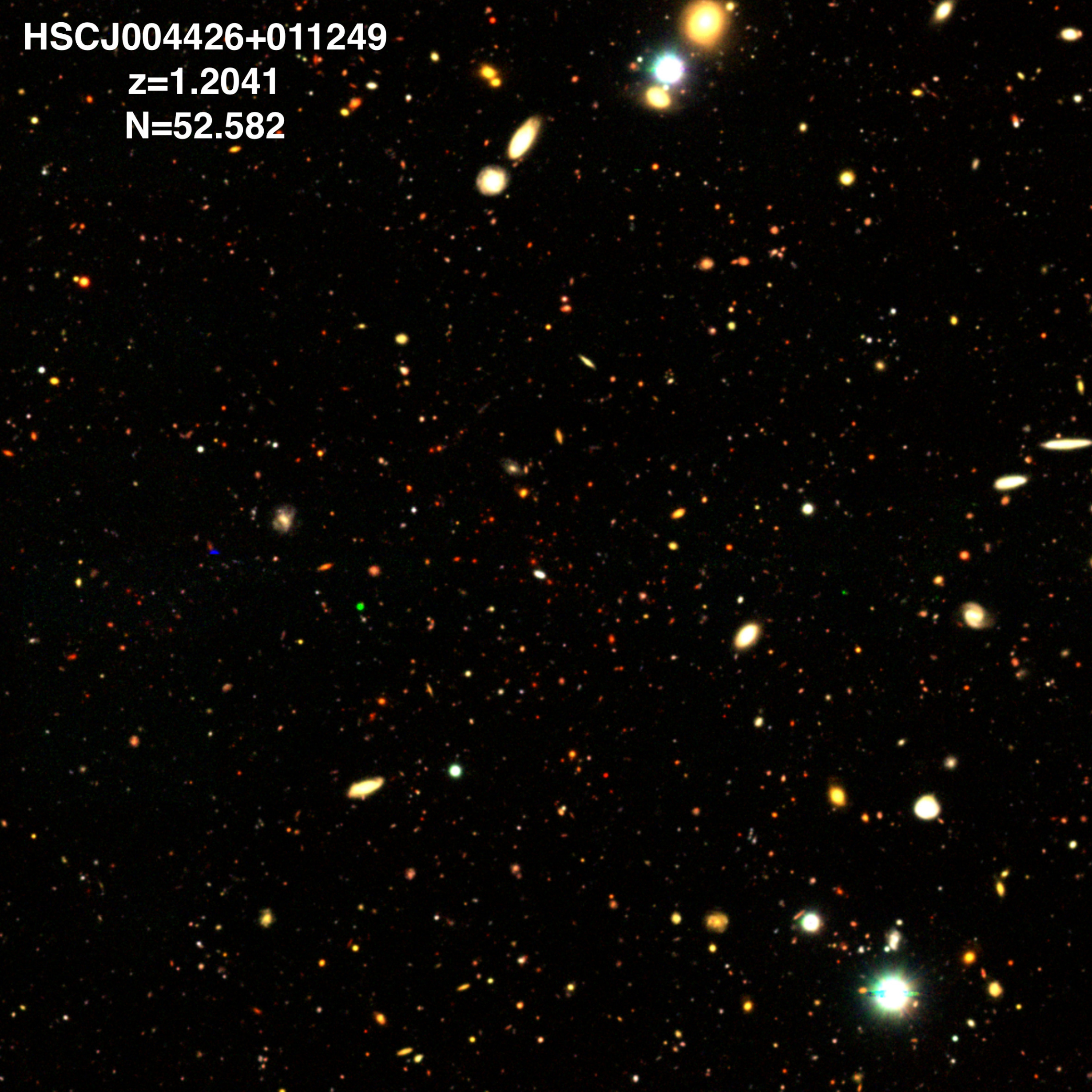} 
 \end{center}
\caption{Examples of $gri$-composite color images of low- ({\it
    left}), mid- ({\it middle}), and high-redshift ({\it right})
  CAMIRA clusters. The size of each image is $\sim 5\farcm5$.
  {Alt text: concentrations of red galaxies are shown.}
}\label{fig:hsc_camira_ex} 
\end{figure*}

We show examples of color images of CAMIRA clusters in
Figure~\ref{fig:hsc_camira_ex}.

\subsection{Comparison with an X-ray cluster catalog}\label{sec:x-ray}

We cross match the CAMIRA cluster catalog with SRG/eROSITA All-Sky
Survey first cluster catalog \citep[hereafter eRASS1;][]{bulbul24}.
The matching condition is the redshift difference of
$|\Delta z|\leq 0.05$ and the physical transverse distance of $R\leq
1h^{-1}$Mpc, where the transverse distance is computed assuming the
redshift of the eRASS1 cluster. There are 269 clusters that are
selected by this matching condition.\footnote{For comparison, if we
adopt a relaxed condition of $|\Delta z|\leq 0.07$ and
$R\leq 2h^{-1}$Mpc, we obtain 315 matched clusters.} We note that less
than half of the HSC-SSP footprint has an overlap with the eRASS1 footprint.

\begin{figure}
 \begin{center}
   \includegraphics[width=8.6cm]{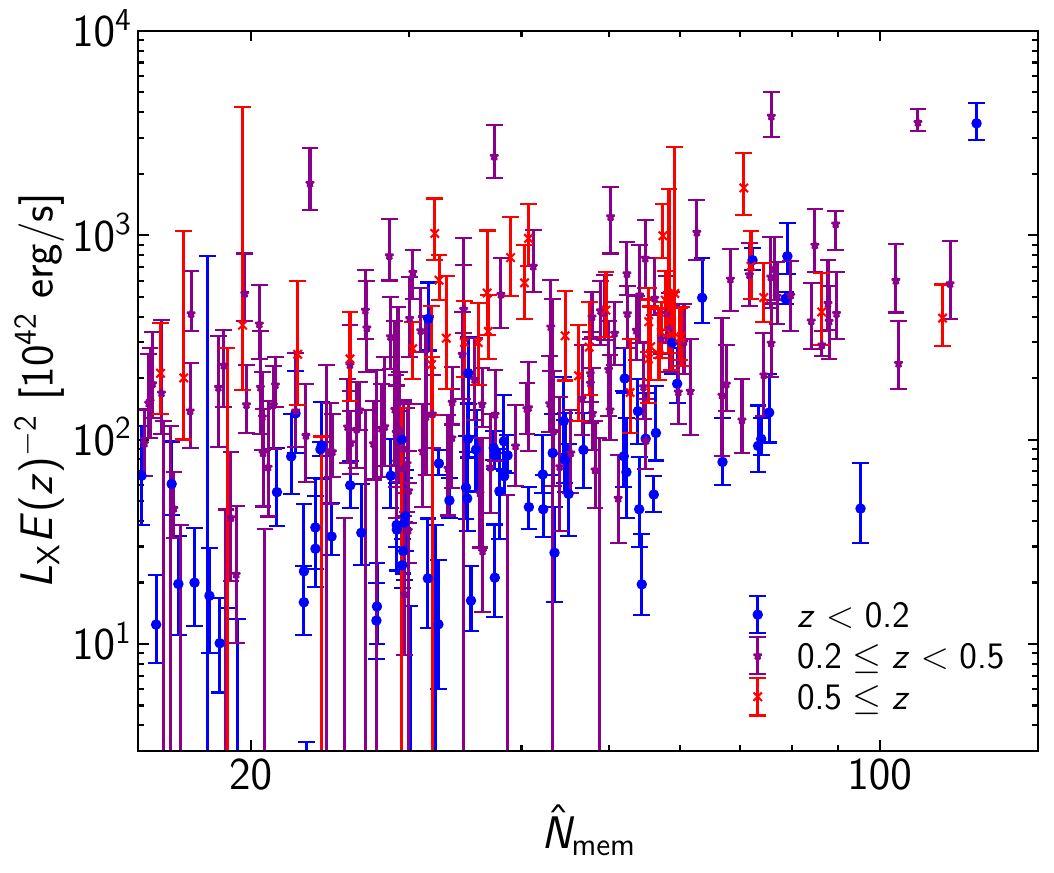} 
 \end{center}
\caption{Comparison between richness $\hat{N}_{\mathrm{mem}}$ and
  eRASS1 X-ray luminosity $L_X$ with the correction of $[E(z)]^{-2}$,
  where $E(z)$ the Hubble parameter normalized by the Hubble
  constant. Here the X-ray luminosity refers to the bolometric
  luminosity within $r_{500}$ \citep[see][]{bulbul24}. Clusters at
  $z<0.2$, $0.2\leq z<0.5$, and $0.5\leq z$ are shown by blue circles,
  magenta stars, and red crosses, respectively.
  {Alt text: the x axis shows the richness from 15 to 200, and the y
    axis shows the X-ray luminosity from three times 10 to 42 to 10 to
    46 erg per second.}
}\label{fig:erass_nlum} 
\end{figure}

For the matched clusters, we compare richness $\hat{N}_{\mathrm{mem}}$
with the bolometric X-ray luminosity $L_X$ in Figure~\ref{fig:erass_nlum}. 
We find a positive correlation between these two quantities, which
is consistent with previous analysis of X-ray properties of CAMIRA
clusters \citep[e.g.,][]{oguri18a,okabe19,willis21,ota23}. The relatively large
scatter mainly reflects the large scatter of the richness-mass
relation. A detailed analysis of the richness-X-ray luminosity
scaling relation requires the correction of the Malmquist bias.

\begin{figure}
 \begin{center}
   \includegraphics[width=8.6cm]{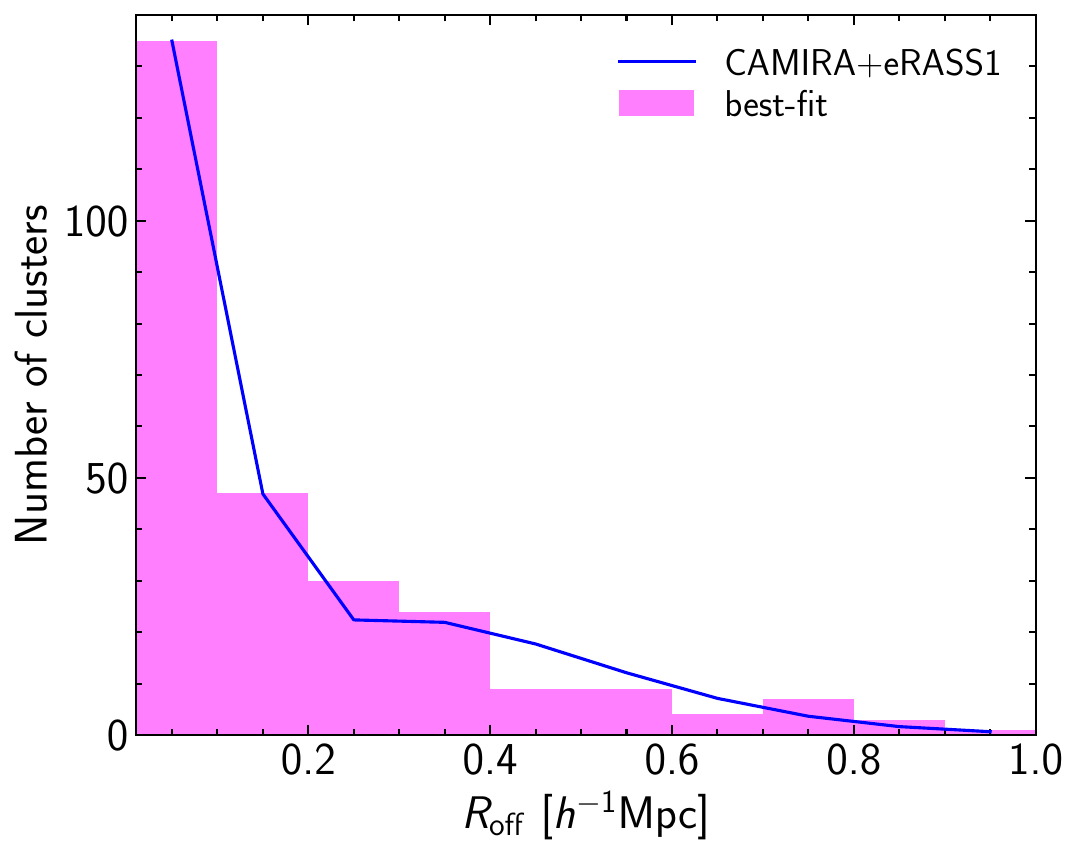} 
 \end{center}
\caption{Histogram of positional offsets $R_{\mathrm{off}}$, which are
  defined by physical transverse distances, between
    CAMIRA cluster centers and eRASS1 X-ray cluster centers. The solid
    line shows the best-fitting model assuming a two-component model.
    {Alt text: The x axis shows the positional offset from 0 to 1
      megaparsec.}
}\label{fig:erass_off} 
\end{figure}

It is known that centers of clusters are often misidentified for
optically selected clusters. Such mis-centering is important e.g.,
when interpreting stacked weak lensing signal around optically
selected clusters. Since X-ray centers trace potential minima of
clusters fairly well, a comparison of optically selected clusters
with X-ray clusters enable us to quantitatively study the mis-centering
effect. Figure~\ref{fig:erass_off} shows the mis-centering distribution
derived from the matched sample. Following \citet{oguri18a}, we fit
the observed histogram with the following two-component model
\begin{align}
p(R_{\mathrm{off}})=&f_{\mathrm{cen}}\frac{R_{\mathrm{off}}}{\sigma_1^2}\exp\left(-\frac{R_{\mathrm{off}}^2}{2\sigma_1^2}\right)
\nonumber\\
&+(1-f_{\mathrm{cen}})\frac{R_{\mathrm{off}}}{\sigma_2^2}\exp\left(-\frac{R_{\mathrm{off}}^2}{2\sigma_2^2}\right),
\label{eq:poff}
\end{align}
where $f_{\mathrm{cen}}$ is the fraction of mis-centered component and
$\sigma_1$ and $\sigma_2$ are standard deviations of centered and
mis-centered components, respectively. We fit all these three
parameters simultaneously to find $f_{\mathrm{cen}}=0.590\pm0.035$,
$\sigma_1=(0.055\pm0.004)h^{-1}$Mpc, and $\sigma_2=(0.292\pm0.016)h^{-1}$Mpc.
The value of $f_{\mathrm{cen}}$ is slightly lower but is consistent
with the result in \citet{oguri18a}. This value is also consistent with
$f_{\mathrm{cen}}\sim 0.6$ obtained from the similar analysis using
X-ray clusters \citep{willis21,ota23,okabe25}. \citet{ding25}
derive the mis-centering distribution of CAMIRA clusters in comparison
with Sunyaev-Zel'dovich clusters to find $f_{\mathrm{cen}}\sim 0.75$,
significantly higher than the value derived here, although we note that
the higher $\sigma_1$ of $\sigma_1\simeq 0.1h^{-1}$Mpc in their model
may partly be a cause of the high $f_{\mathrm{cen}}$.

\subsection{Stacked weak lensing analysis}\label{sec:stacked_lensing}

The mass-richness relation of HSC-SSP CAMIRA clusters has been
constrained with stacked weak lensing analysis based
on the earlier version of the HSC-SSP data \citep{murata19}. Here we
revisit the weak lensing analysis of the mass-richness relation for
the HSC-SSP final year CAMIRA clusters, using the three-year HSC-SSP
shear catalog \citep{li2022} that covers $\sim 430$~deg$^2$ of the
HSC-SSP footprint with the raw number density of $22.9$~arcmin$^{-2}$.

We derive differential surface mass density profiles $\Delta\Sigma(R)$
for subsamples of CAMIRA clusters following \citet{medezinski18}.
Errors on the measurements include intrinsic ellipticities of
galaxies. We use the $P(z)$ cut method, for which an integrated
probability of the photometric redshift probability distribution
function (PDF) above redshift $z_{\mathrm{cl}}+\Delta z$ exceeds a
threshold $p_{\mathrm{cut}}$ is used for the weak lensing analysis, to
construct the background galaxy sample. Throughout the paper, we adopt
DNNz photometric redshifts (A. J. Nishizawa et al., in prep.) for the
photometric redshifts of source galaxies, $\Delta z=0.1$,
and $p_{\mathrm{cut}}=0.98$. While we derive $\Delta\Sigma(R)$ in the
range of the physical radius between $0.1$ and $10h^{-1}\mathrm{Mpc}$
with a logarithmic bin width of $\Delta(\log r)=0.1$, we use
$\Delta\Sigma(R)$ only in the range
$0.3h^{-1}\mathrm{Mpc}<r<5h^{-1}\mathrm{Mpc}$ in the
subsequent analysis. The removal of the small-scale signal at
$r<0.3h^{-1}\mathrm{Mpc}$ mitigates the impact of the cluster member
dilution effect \citep[see][for detailed discussions]{medezinski18},
whereas we remove the large-scale signals at $r>5h^{-1}\mathrm{Mpc}$
where the impact of the cosmic shear on the covariance of
$\Delta\Sigma(R)$ becomes more important
\citep{hoekstra03,dodelson04}, meaning that the current errors from the
intrinsic shapes of galaxies are underestimated at those radii.

We divide CAMIRA clusters into four redshift bins
($0.2<z_{\mathrm{cl}}<0.4$, $0.4<z_{\mathrm{cl}}<0.6$,
$0.6<z_{\mathrm{cl}}<0.8$, $0.8<z_{\mathrm{cl}}<1.0$)
and four richness bins ($15<\hat{N}_{\mathrm{mem}}<20$,
$20<\hat{N}_{\mathrm{mem}}<30$, $30<\hat{N}_{\mathrm{mem}}<50$, 
$50<\hat{N}_{\mathrm{mem}}<200$), and measure $\Delta\Sigma(R)$ for
each of the 16 subsamples. Each subsample contains $\sim 1000-1500$
clusters for the lowest redshift bin $0.2<z_{\mathrm{cl}}<0.4$,
and $\sim 10-100$ clusters at the highest redshift bin
$0.8<z_{\mathrm{cl}}<1.0$. We fit the signals with a model of
the richness-mass relation considered in \citet{murata19}. 
Specifically, we assume the following conditional PDF of the richness
$N=\hat{N}_{\mathrm{mem}}$ for a halo with mass
$M=M_{200{\mathrm{m}}}$ (the mass inside a radius within which the
overdensity is 200 times the mean matter density of the Universe) and
redshift $z=z_{\mathrm{cl}}$  
\begin{equation}
  P(\ln N | M,\,z)=\frac{1}{\sqrt{2\pi}\sigma_{\ln M|M,z}}\exp\left[-\frac{x^2(N,\,M,\,z)}{2\sigma_{\ln M|M,z}^2}\right],
  \label{eq:nm_def}
\end{equation}
\begin{align}
  x(N,\,M,\,z)=&\ln N -
  \left[A+B\ln\left(\frac{M}{M_0}\right)+B_z\ln\left(\frac{1+z}{1+z_0}\right)\right.\nonumber\\
    &+C_z\left.\left\{\ln\left(\frac{1+z}{1+z_0}\right)\right\}^2\right],
  \label{eq:nm_x_def}
\end{align}
with the pivot mass $M_0=3\times 10^{14}h^{-1}M_\odot$ and the pivot
redshift $z_0=0.6$. The scatter $\sigma_{\ln M|M,z}$ also depends on
the mass and redshift as
\begin{align}
  \sigma_{\ln M|M,z}=& \sigma_0+q\ln\left(\frac{M}{M_0}\right)+q_z\ln\left(\frac{1+z}{1+z_0}\right)\nonumber\\
    &+p_z\left\{\ln\left(\frac{1+z}{1+z_0}\right)\right\}^2.
  \label{eq:nm_sigma_def}
\end{align}
Given richness-mass relation, we compute the average surface mass
density profile for each richness bin $\alpha$ and redshift
bin $\beta$ as
\begin{align}
  \Delta\Sigma_{\alpha,\,\beta}(R)=&\frac{1}{\bar{n}_{\alpha,\,\beta}}\int_{M_{\mathrm{min}}}^{M_{\mathrm{max}}}dM\frac{dn}{dM}S(M,\,\bar{z}_\beta|N_{\alpha,\mathrm{min}},\,N_{\alpha,\mathrm{max}})\nonumber\\
  &\times\frac{\Delta\Sigma(R;\,M,\,\bar{z}_\beta)}{1-\langle\Sigma_{\mathrm{cr}}^{-1}\rangle_{\alpha,\,\beta}\Sigma(R;\,M,\,\bar{z}_\beta)},
\end{align}
\begin{equation}
 \bar{n}_{\alpha,\,\beta}=\int_{M_{\mathrm{min}}}^{M_{\mathrm{max}}}dM\frac{dn}{dM}S(M,\,\bar{z}_\beta|N_{\alpha,\mathrm{min}},\,N_{\alpha,\mathrm{max}}),
\end{equation}
\begin{align}
  S(M,\,z|N_{\alpha,\mathrm{min}},\,N_{\alpha,\mathrm{max}})=&\frac{1}{2}
  \left[\mathrm{erf}\left(\frac{x(N_{\alpha,\mathrm{max}},\,M,\,z)}{\sqrt{2}\sigma_{\ln M|M,z}}\right)\right.\nonumber\\
    &\left.-\mathrm{erf}\left(\frac{x(N_{\alpha,\mathrm{min}},\,M,\,z)}{\sqrt{2}\sigma_{\ln M|M,z}}\right)\right],
\end{align}
where $\Sigma_{\mathrm{cr}}$ is the critical surface density,
$\Sigma(R)$ is the surface mass density profile,
$\mathrm{erf}(x)$ is the error function, $\bar{z}_\beta$ is the center
of the redshift bin $\beta$, and we adopt $M_{\mathrm{min}}=10^{12}h^{-1}M_\odot$ and
$M_{\mathrm{max}}=10^{16}h^{-1}M_\odot$.  We use the \citet{tinker08} mass function
for $dn/dM$. For computing $\Delta\Sigma(R)$ and $\Sigma(R)$, we
assume the Baltz-Marshall-Oguri \citep[BMO;][]{baltz09,oguri11}
density profile taking account of the mis-centering effect, and include
the so-called two-halo term. A more detailed description of the modeling
of $\Delta\Sigma(R)$ and $\Sigma(R)$ in this paper is given in
Appendix~\ref{app:halo_lensing}.

\begin{table}
  \tbl{Model parameters, priors, and values estimated from the lensing
    profile analysis.\footnotemark[$*$] }{%
  \begin{tabular}{ccc}
      \hline
      Parameter & Prior & Median and Error\\ 
      \hline
      $f_{\mathrm{cen}}$ & [$0.5$, $0.7$] & $0.566_{-0.041}^{+0.045}$\\
      $\sigma_{\mathrm{off}}$ & [$0.2$, $0.4$] & $0.249_{-0.020}^{+0.025}$\\
      $A$ & [$2$, $5$] & $3.671_{-0.112}^{+0.058}$\\
      $B$ & [$0$, $2$] & $0.756_{-0.047}^{+0.040}$\\
      $B_z$ & [$-10$, $10$] & $-0.209_{-0.341}^{+0.330}$\\
      $C_z$ & [$-10$, $10$] & $-3.025_{-2.221}^{+2.101}$\\
      $\sigma_0$ & [$0$, $1.5$] & $0.136_{-0.069}^{+0.083}$\\
      $q$ & [$-1$, $1$]& $-0.008_{-0.022}^{+0.014}$\\
      $q_z$ & [$-10$, $10$] & $-0.333_{-0.251}^{+0.291}$\\
      $p_z$ & [$-10$, $10$] & $2.471_{-1.575}^{+1.732}$\\
      \hline 
    \end{tabular}}\label{tab:mcmc}
\begin{tabnote}
\footnotemark[$*$] All the priors are a flat prior in the range
indicate by [$p_{\mathrm{min}}$, $p_{\mathrm{max}}$]. The
mis-centering parameter  $\sigma_{\mathrm{off}}$ is in units of
$h^{-1}$Mpc.
\end{tabnote}
\end{table}

\begin{figure*}
 \begin{center}
   \includegraphics[width=16.8cm]{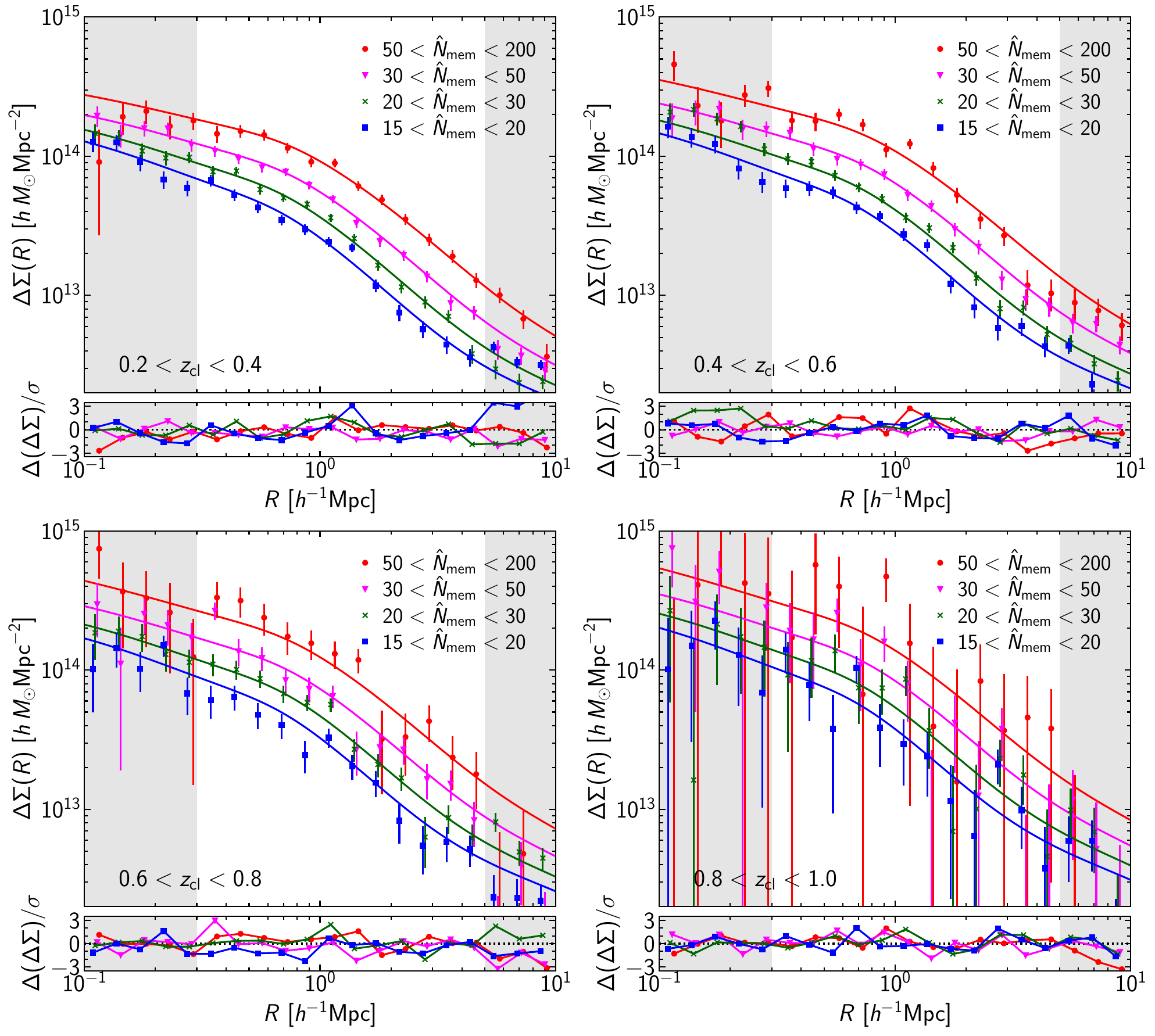} 
 \end{center}
\caption{Lensing profiles measured from the three-year HSC-SSP shear
  catalog. Different panels show results for different redshift
  bins. In each panel, we plot profiles for different richness bins
  with different symbols. Shaded regions indicate ranges of the radii
  that are excluded from the analysis (see text for more
  details). Solid lines indicate model predictions with the best-fit
  parameters listed in Table~\ref{tab:mcmc}. In sub-panels, we
    also show residuals (i.e., observed minus model predicted values)
    normalized by errors. 
    {Alt text: In each panel, the x axis shows the distance from the
      cluster center from 0.1 to 10 megaparsec. The y axis shows the
      differential surface mass distribution.}
}\label{fig:camira_lensing} 
\end{figure*}

We use {\tt emcee} \citep{foreman13} for the Markov chain Monte Carlo
(MCMC) analysis. We assume a flat Universe with the matter density
parameter $\Omega_{\mathrm{m}}=0.3$, the cosmological constant
$\Omega_\Lambda=0.7$, the baryon density parameter $\Omega_{\mathrm{b}}=0.05$,
the dimensionless Hubble constant $h=0.7$, the spectral index
$n_{\mathrm{s}}=0.96$, and the normalization of the matter power spectrum
$\sigma_8=0.81$, which are also assumed when deriving stacked
differential surface density profiles from the three-year HSC-SSP
shear catalog. Model parameters include $f_{\mathrm{cen}}$ and
$\sigma_{\mathrm{off}}$ from the mis-centering effect and eight parameters
in equations~(\ref{eq:nm_x_def}) and (\ref{eq:nm_sigma_def}) to model
the richness-mass relation. We ignore the redshift and mass dependence
of the mis-centering parameters, and assume flat priors in relatively
narrow parameter range that is consistent with the result of
cross-matching with X-ray clusters presented in Sec.~\ref{sec:x-ray}.
The model parameters and their prior ranges are summarized in
Table~\ref{tab:mcmc}. 
 
From the MCMC analysis, we find that our model fits the observed
lensing profiles reasonably well with the minimum $\chi^2$ of
$\chi_{\mathrm{min}}/\mathrm{dof}=182.4/182$.
In Table~\ref{tab:mcmc}, we summarize the median and 68\% credible
interval of each model parameter after marginalizing over the other
parameters. We find that model parameters of the richness-mass
relation are broadly consistent with those in \citet{murata19}. We
emphasize that, aside from using the updated cluster and weak lensing
shear catalogs, our analysis differs from \citet{murata19} in that we
only use lensing constraints, while \citet{murata19} combine lensing
and abundance constraints. By adding constraints from the abundance,
constraints on model parameters can be improved further, although
in that case a degeneracy between the scatter of the richness-mass
relation and cosmological parameters (especially $\sigma_8$) have to
be taken into account as emphasized in \citet{murata19}. We compare
observed lensing profiles with our best-fit model in
Figure~\ref{fig:camira_lensing}. 

\begin{figure}
 \begin{center}
   \includegraphics[width=8.0cm]{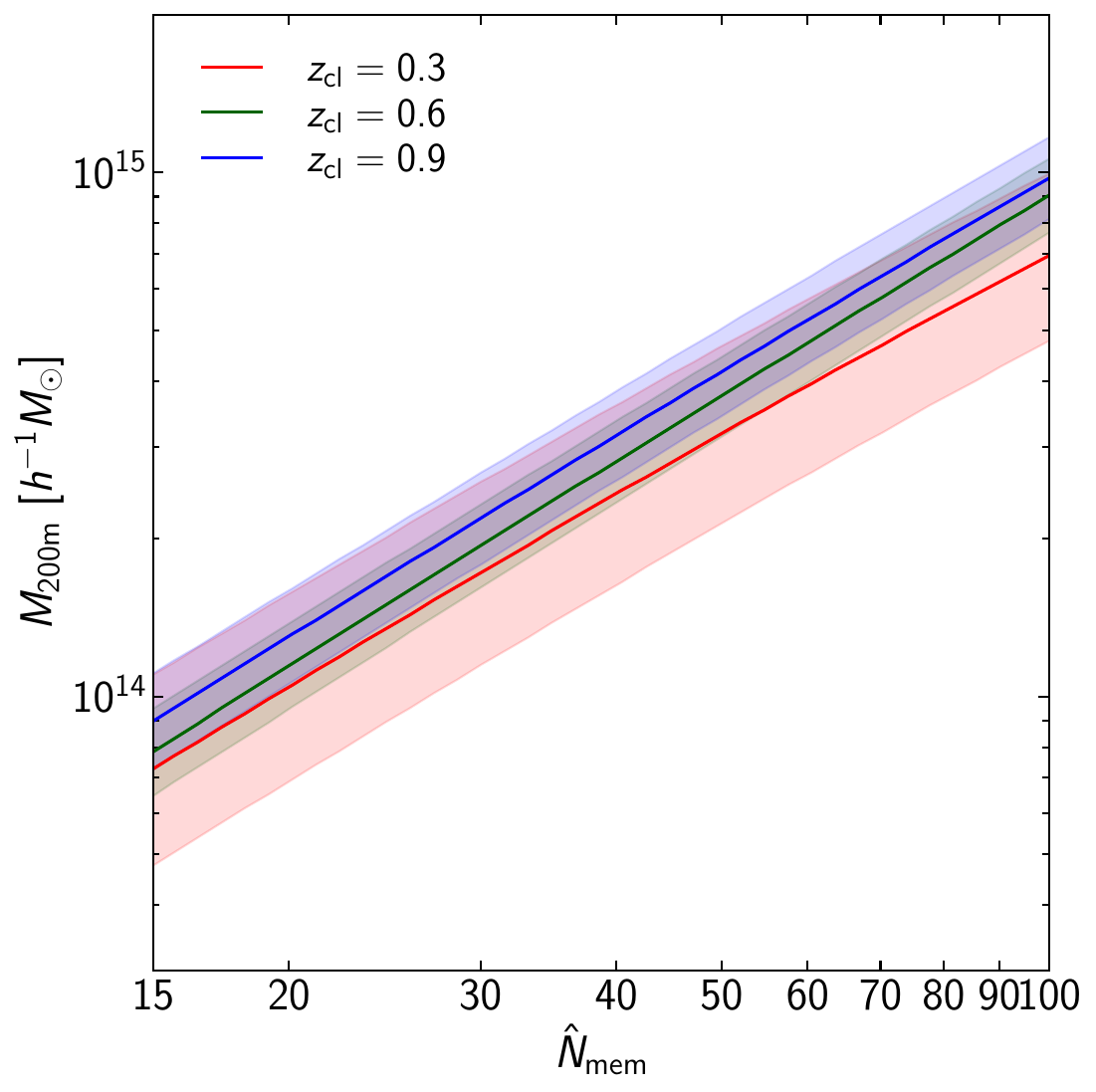} 
 \end{center}
\caption{The mass-richness relations, obtained from
  equation~(\ref{eq:mn_def}), at $z=0.3$, $0.6$, and $0.9$ for our
  best-fit model parameters listed in Table~\ref{tab:mcmc}. Solid
  lines and shaded regions show medians and the 16th and 84th
  percentiles of the PDF of the halo mass for a fixed richness. 
  {Alt text: The x axis shows the richness from 15 to 100, and the y
    axis shows from 3 times 10 to 13 to 2 times 10 to 15 solar masses.}
}\label{fig:camira_nm} 
\end{figure}

Given the richness-mass relation constrained from observed lensing
profiles, we can derive the mass-richness relation using the Bayes
theorem \citep{murata19}. 
\begin{equation}
P(\ln M|N,\,z)=\frac{P(\ln N|M,\,z)P(\ln M|z)}{\int d\ln M P(\ln  N|M,\,z)P(\ln M|z)},
\label{eq:mn_def}
\end{equation}
where $P(\ln N|M,\,z)$ is defined in equation~(\ref{eq:nm_def}) and we
use the halo mass function
\begin{equation}
P(\ln M|z)=\frac{dn}{d\ln M},
\end{equation}
for the PDF of the halo mass. Figure~\ref{fig:camira_nm} shows examples
of the mass-richness relation at three redshifts for our best-fit
model parameters. We find that the richness of
$\hat{N}_{\mathrm{mem}}=20$ corresponds to the halo mass of
$M_{200{\mathrm{m}}}\sim 10^{14}h^{-1}M_\odot$. The larger scatter at
lower redshift is presumably due to the lack of $u$-band as well
as saturations of fluxes of bright galaxies in HSC-SSP.

Once the mass-richness relation is well constrained, one can in
principle use the observed abundance of clusters to place constraints
on cosmological parameters, in particular $\Omega_{\mathrm{m}}$ and $\sigma_8$.
However, in order to place accurate constraints on cosmological
parameters, the projection effect arising from the asphericity of
clusters as well as neighboring galaxies have to be carefully taken
into account \citep[e.g.,][]{osato18,sunayama20,wu22,lee25}.
For instance, \citet{sunayama20} show that optically-selected
  clusters can exhibit lensing signals that deviate from expectations
  based on a statistically isotropic halo model by up to a factor of
  1.2 due to the inclusion of clusters embedded within filaments
  aligned with the line-of-sight direction in the optically selected
  cluster sample.
Furthermore, various additional systematic errors have also to be
carefully taken into account, including photometric redshift errors,
shear measurement biases, additional contributions to the error
covariance matrix, the cosmology dependence of the observed lensing
profiles, and theoretical modeling uncertainties. We leave such
exploration to future work. 

\section{Photometric LRG catalogs}\label{sec:lrg}


From the HSC-SSP final year dataset, we also construct photometric LRG 
catalogs in the photometric redshift range of
$0.05\leq z_{\mathrm{LRG}}\leq 1.25$ and the stellar mass range of
$M_*\geq 10^{10.3}M_\odot$ using the method 
described in Sec.~\ref{sec:algorithm}.\footnote{While they are not
necessarily luminous with such low stellar mass limit, we keep the
name ``photometric LRG'' for historical reasons.} The photometric LRG
catalogs  contain 6705294 galaxies from the Wide layer after applying
for the bright star mask, and 119955 galaxies from the Deep+UD layer after
applying for the bright star mask. We note that the photometric LRG
catalogs are constructed independently of the CAMIRA cluster catalogs
presented in Sec.~\ref{sec:cluster} and photometric LRGs are not used
for cluster finding, although they share the same stellar population
synthesis model calibrated by spectroscopic redshifts.  

\begin{figure}
 \begin{center}
   \includegraphics[width=8.6cm]{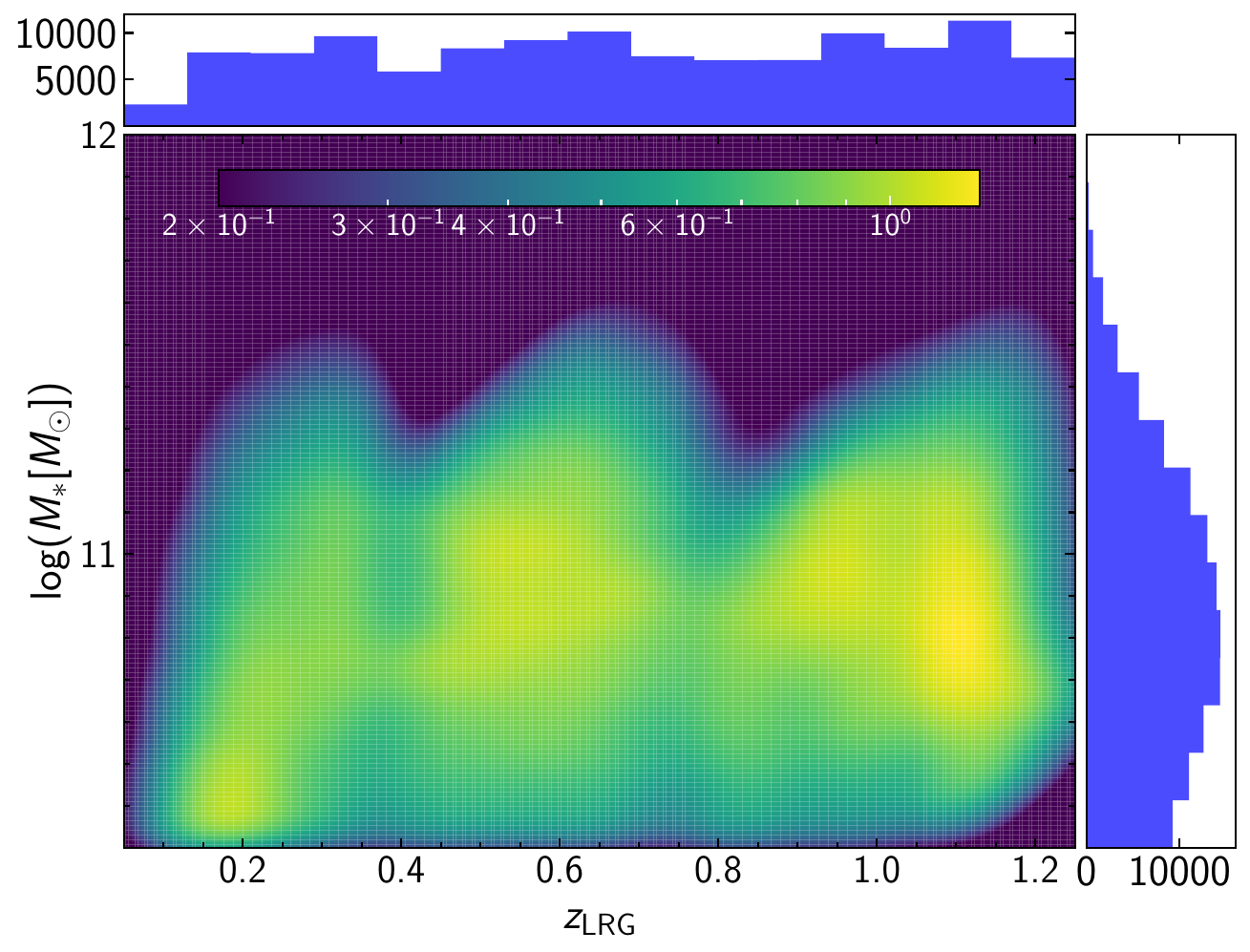} 
 \end{center}
\caption{The color density map of photometric LRGs in the
  $z_{\mathrm{LRG}}$--$M_*$ plane, where $z_{\mathrm{LRG}}$ and $M_*$
  refer to photometric redshifts and stellar masses of LRGs,
  respectively. The density is calculated by a kernel density
  estimation using Gaussian kernels. Due to the large number of
  photometric LRGs for Wide, here we show the density map for Deep+UD
  as they are similar between Wide and Deep+UD.
  Top and right panels show the histograms of $z_{\mathrm{LRG}}$ and
  $M_*$, respectively. 
  {Alt text: The x axis shows the redshift from 0.1 to 1.25 and the y
    axis shows the logarithm of stellar mass in solar masses from 10.3
    to 12.} 
}\label{fig:lrg_mz}  
\end{figure}

Figure~\ref{fig:lrg_mz} shows the density map in the redshift--stellar
mass plane. It is found that the catalog contains many photometric
LRGs out to $z\sim 1.2$. In the density map, there are features at
$z_{\mathrm{LRG}}\sim 0.4$ and $\sim 0.8$, which reflects that fact
that the 4000{\AA} break of the spectral energy distribution of an LRG
moves from $g-r$ to $r-i$ and from $r-i$ to $i-z$ at these redshifts
\citep[e.g., see also][]{rykoff14}. The Figure also suggests that the
photometric LRG sample is nearly complete out to
$z_{\mathrm{LRG}}\sim 1.1$, while at the highest redshift
$z_{\mathrm{LRG}}\sim 1.2$ the LRG sample is incomplete in the sense
that LRGs with low stellar masses around the limit $M_*\geq
10^{10.3}M_\odot$ are not included due to the magnitude limit of the
HSC-SSP.

\begin{figure}
 \begin{center}
   \includegraphics[width=8.6cm]{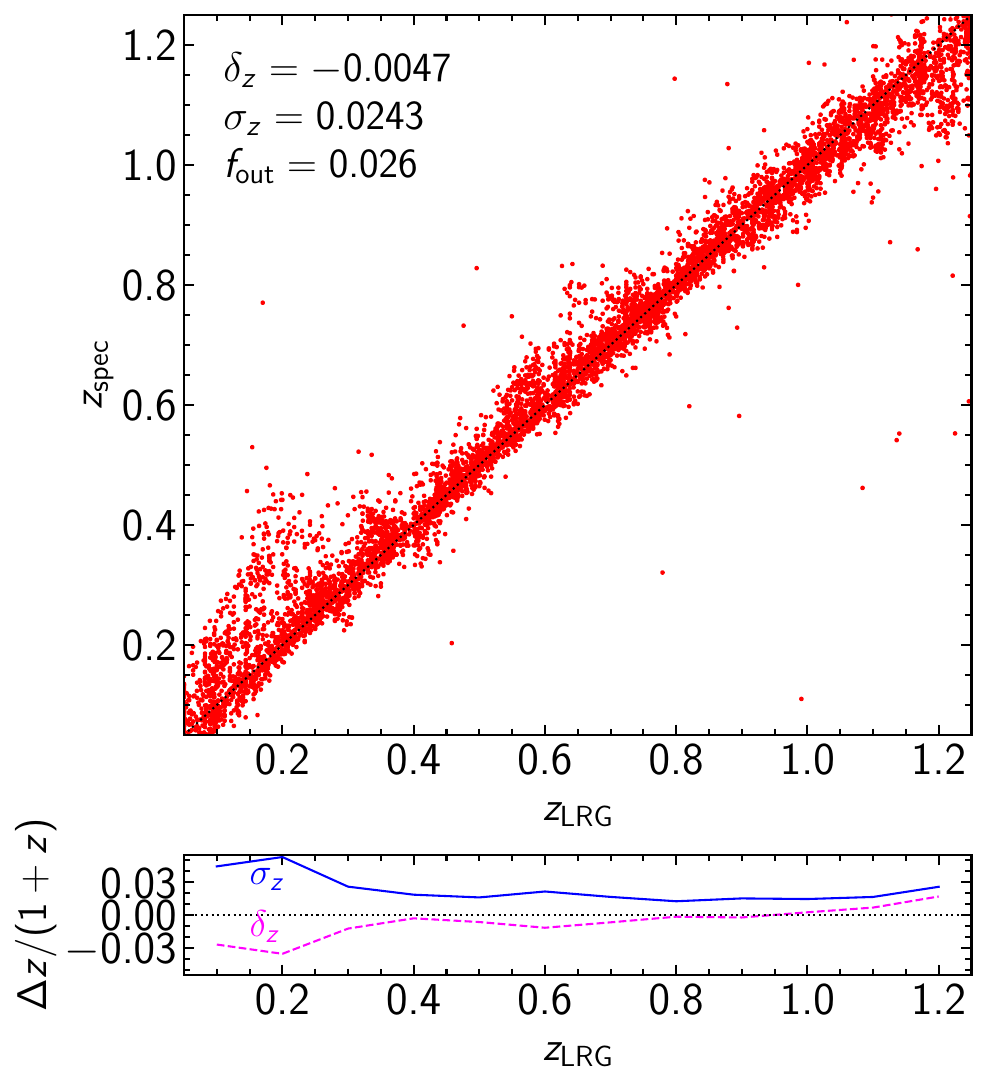} 
 \end{center}
\caption{Similar to Figure~\ref{fig:camira_zcomp}, but for the
  comparison of photometric redshifts and spectroscopic redshifts of
  photometric LRGs. Among all 590519 photometric LRGs that are matched
  with spectroscopic galaxies in the Wide layer, we select a subsample
  such that roughly 300 photometric LRGs in each redshift bin with the
  width of $0.05$ are randomly selected.  The comparison is made for
  this subsample of 7201 galaxies. The outlier fraction is
  $f_{\mathrm{out}}=0.026$ for the whole subsample.
  {Alt text: the x and y axes show the redshift from 0.1  to 1.25.}
}\label{fig:lrg_zcomp} 
\end{figure}

To check the photometric redshift accuracy and precision, we cross
match the photometric redshift catalog with the spectroscopic catalog
described in Sec.~\ref{sec:spec_z}. For the Wide photometric LRG
catalog, we find 590519 photometric LRGs have spectroscopic redshifts. 
The comparison for a subsample of photometric LRGs with spectroscopic
redshifts shown in Figure~\ref{fig:lrg_zcomp} indicates that the
photometric redshifts are accurate and precise for a wide range in the
redshift. We find the relatively large bias and scatter at
$z_{\mathrm{LRG}}\lesssim 0.3$, which is partly due to the lack of
$u$-band as well as saturations of fluxes of bright galaxies in
HSC-SSP. We refer to \citet{2019MNRAS.489.5202S} for a detailed
  exploration on how adding $u$-band photometry to the HSC-SSP data
  improves photometric redshifts across wide range of redshifts.
We note that, due to combining photometric redshifts of many member
galaxies that effectively remove outliers, cluster photometric redshifts are less
degraded at $z_{\mathrm{cl}}\lesssim 0.3$ (see Figure~\ref{fig:camira_zcomp}).
The photometric redshift accuracy and precision is stable at
$0.4 \lesssim z_{\mathrm{LRG}} \lesssim 1$ with the scatter of of the
redshift residual
$(z_{\mathrm{LRG}}-z_{\mathrm{spec}})/(1+z_{\mathrm{spec}})$ of $\sigma_z\lesssim 0.02$.
The scatter and the outlier fraction are larger than those for cluster
photometric redshifts (see Figure~\ref{fig:camira_zcomp}), because we
effectively take the average of multiple photometric redshifts of
galaxies to derive cluster photometric redshifts. We also see the sign
of the degradation of the photometric redshift precision and accuracy
at very high redshift, $z_{\mathrm{LRG}} \gtrsim 1.1$.

The new photometric LRG catalog presented in this paper is useful
  in several ways, including the calibration of photometric redshifts
  of galaxies used for the weak lensing analysis
  \citep{2023MNRAS.524.5109R} and the study of the redshift evolution
  of quiescent galaxies over a wide redshift range \citep{ishikawa21}.
 The new photometric LRG catalog is also useful for a detailed study
 of the central density profile of quiescent galaxies with stacked
  weak gravitational lensing \citep{2025arXiv251209342F}.

\section{Summary}\label{sec:summary}

We have presented catalogs of optically-selected clusters and
photometric LRGs that are constructed from the HSC-SSP final year
dataset covering $\sim 1200$~deg$^2$. The cluster catalogs contain
more than 10000 clusters with richness larger than $15$ out to the
redshift of $1.38$. The photometric redshift is accurate and precise
with the scatter $\sigma_z\lesssim 0.01$ out to $z\sim 1$. We derive
lensing profiles of our cluster sample for a wide range in the
richness and redshift to constrain the mass-richness relation.
The photometric LRG catalogs, which are a byproduct of the CAMIRA
cluster finding algorithm, contain more than six million red galaxies
with the stellar mass larger than $10^{10.3}M_*$ out to the redshift
of $1.25$. The photometric redshifts of LRGs are accurate and precise
particularly in the range of $0.4\lesssim z\lesssim 1$ where the
scatter is $\sigma_z\lesssim 0.02$. The catalogs are useful for
various applications in cosmology and astrophysics. 

\section*{Supplementary data} 

The following supplementary data is available at PASJ online.

Supplementary Tables 1, 2, 3, and 4.

\begin{ack}
  We thank an anonymous referee for useful suggestions.
This work was supported by JSPS KAKENHI Grant Numbers JP25H00662, JP25H00672, JP25H01551, JP25K01026, JP25K01044, JP24H00002, JP24K00684, JP23H00108, JP23K22537, JP22K21349.
YTL acknowledge support from National Science and Technology Council of Taiwan under grant NSTC 114-2112-M-001-034.
NO acknowledges partial support from the Organization for the Promotion of Gender Equality at Nara Women’s University.
INC acknowledges the support from the National Science and Technology Council in Taiwan (Grant NSTC 111-2112-M-006-037-MY3 and 114-2112-M-006-017-MY3).
LL thanks support by the National Science and Technology Council of Taiwan under grant  NSTC 114-2112-M-001-041-MY3.

The Hyper Suprime-Cam (HSC) collaboration includes the astronomical communities of Japan and Taiwan, and Princeton University. The HSC instrumentation and software were developed by the National Astronomical Observatory of Japan (NAOJ), the Kavli Institute for the Physics and Mathematics of the Universe (Kavli IPMU), the University of Tokyo, the High Energy Accelerator Research Organization (KEK), the Academia Sinica Institute for Astronomy and Astrophysics in Taiwan (ASIAA), and Princeton University. Funding was contributed by the FIRST program from the Japanese Cabinet Office, the Ministry of Education, Culture, Sports, Science and Technology (MEXT), the Japan Society for the Promotion of Science (JSPS), Japan Science and Technology Agency (JST), the Toray Science Foundation, NAOJ, Kavli IPMU, KEK, ASIAA, and Princeton University. 

This paper makes use of software developed for the Large Synoptic Survey Telescope. We thank the LSST Project for making their code available as free software at  http://dm.lsst.org

This paper is based on data collected at the Subaru Telescope and retrieved from the HSC data archive system, which is operated by the Subaru Telescope and Astronomy Data Center (ADC) at National Astronomical Observatory of Japan. Data analysis was in part carried out with the cooperation of Center for Computational Astrophysics (CfCA), National Astronomical Observatory of Japan. The Subaru Telescope is honored and grateful for the opportunity of observing the Universe from Maunakea, which has the cultural, historical and natural significance in Hawaii. 
\end{ack}


\section*{Data availability} 
 The data underlying this article will be made available at {\tt https://hsc.mtk.nao.ac.jp/ssp/}.

\appendix 
\section{SQL query for an input galaxy catalog}\label{app:sql_query}

Here we show an example of the SQL query to extract an input galaxy
catalog from the HSC database, for which we run the CAMIRA algorithm.

\begin{footnotesize}
\begin{verbatim}
SELECT photo.ra, photo.dec, 
       afb.g_undeblended_convolvedflux_3_15_mag
          - afb.z_undeblended_convolvedflux_3_15_mag
          + photo.z_cmodel_mag - photo.a_g
          - offsets.g_mag_offset as gmag, 
       photo.g_cmodel_magerr as gmag_err,
       afb.r_undeblended_convolvedflux_3_15_mag
          - afb.z_undeblended_convolvedflux_3_15_mag
          + photo.z_cmodel_mag - photo.a_r
          - offsets.r_mag_offset as rmag, 
       photo.r_cmodel_magerr as rmag_err,
       afb.i_undeblended_convolvedflux_3_15_mag
          - afb.z_undeblended_convolvedflux_3_15_mag
          + photo.z_cmodel_mag - photo.a_i
          - offsets.i_mag_offset as imag, 
       photo.i_cmodel_magerr as imag_err,
       photo.z_cmodel_mag - photo.a_z
          - offsets.z_mag_offset as zmag, 
       photo.z_cmodel_magerr as zmag_err,
       afb.y_undeblended_convolvedflux_3_15_mag
           - afb.z_undeblended_convolvedflux_3_15_mag
           + photo.z_cmodel_mag - photo.a_y
           - offsets.y_mag_offset as ymag, 
       photo.y_cmodel_magerr as ymag_err,
       photo.object_id,
       (mask.g_mask_brightstar_halo
           OR mask.r_mask_brightstar_halo
           OR mask.i_mask_brightstar_halo
           OR mask.z_mask_brightstar_halo
           OR mask.y_mask_brightstar_halo
           OR mask.g_mask_brightstar_ghost
           OR mask.r_mask_brightstar_ghost
           OR mask.i_mask_brightstar_ghost
           OR mask.z_mask_brightstar_ghost
           OR mask.y_mask_brightstar_ghost
           OR mask.g_mask_brightstar_blooming
           OR mask.r_mask_brightstar_blooming
           OR mask.i_mask_brightstar_blooming
           OR mask.z_mask_brightstar_blooming
           OR mask.y_mask_brightstar_blooming)
           as mask_bstar_combined
       
FROM s23b_wide.forced as photo
     LEFT JOIN s23b_wide.stellar_sequence_offsets
        as offsets USING (skymap_id)
     LEFT JOIN s23b_wide.forced5 as afb USING (object_id)
     LEFT JOIN s23b_wide.meas as meas USING (object_id)
     LEFT JOIN s23b_wide.masks as mask USING (object_id)
     
WHERE  photo.g_pixelflags_edge is False
  AND  photo.r_pixelflags_edge is False
  AND  photo.i_pixelflags_edge is False
  AND  photo.z_pixelflags_edge is False
  AND  photo.y_pixelflags_edge is False

  AND  photo.g_pixelflags_interpolatedcenter is False
  AND  photo.r_pixelflags_interpolatedcenter is False
  AND  photo.i_pixelflags_interpolatedcenter is False
  AND  photo.z_pixelflags_interpolatedcenter is False
  AND  photo.y_pixelflags_interpolatedcenter is False

  AND  photo.g_pixelflags_crcenter is False
  AND  photo.r_pixelflags_crcenter is False
  AND  photo.i_pixelflags_crcenter is False
  AND  photo.z_pixelflags_crcenter is False
  AND  photo.y_pixelflags_crcenter is False

  AND  photo.g_cmodel_flag is False
  AND  photo.r_cmodel_flag is False
  AND  photo.i_cmodel_flag is False
  AND  photo.z_cmodel_flag is False
  AND  photo.y_cmodel_flag is False

  AND  afb.g_undeblended_convolvedflux_3_15_flag is False
  AND  afb.r_undeblended_convolvedflux_3_15_flag is False
  AND  afb.i_undeblended_convolvedflux_3_15_flag is False
  AND  afb.z_undeblended_convolvedflux_3_15_flag is False
  AND  afb.y_undeblended_convolvedflux_3_15_flag is False

  AND  photo.isprimary is True

  AND  meas.g_inputcount_value > 1
  AND  meas.r_inputcount_value > 1
  AND  meas.i_inputcount_value > 2
  AND  meas.z_inputcount_value > 2
  AND  meas.y_inputcount_value > 2

  AND  (afb.r_undeblended_convolvedflux_3_15_mag
         - afb.z_undeblended_convolvedflux_3_15_mag
         + photo.z_cmodel_mag - photo.a_r
         - offsets.r_mag_offset) < 28.0
  AND  (afb.i_undeblended_convolvedflux_3_15_mag
         - afb.z_undeblended_convolvedflux_3_15_mag
         + photo.z_cmodel_mag - photo.a_i
         - offsets.i_mag_offset) < 28.0
  AND  (photo.z_cmodel_mag - photo.a_z
         - offsets.z_mag_offset) < 24.0
  AND  photo.z_cmodel_magerr < 0.1
  AND  photo.i_extendedness_value > 0.9 
\end{verbatim}
\end{footnotesize}

\section{Color cuts of spectroscopic galaxies for the calibration of
  the red-sequence}\label{app:color_cuts}

Here we summarize color cuts for spectroscopic galaxies for
calibrating the red-sequence, which are updated from the ones
presented in \citet{oguri18a}. We adopt the following color cuts
\begin{equation}
  g-r>
\left\{
    \begin{array}{l}
      0.398+2.9z_{\mathrm{sepc}}\;\;\;(z_{\mathrm{sepc}} \leq 0.38), \\
      1.861-0.95z_{\mathrm{sepc}}\;\;\;(0.38 < z_{\mathrm{sepc}}),
    \end{array}
  \right.
\label{eq:gr1}
\end{equation}
\begin{equation}
  r-i>
\left\{
    \begin{array}{l}
      0.2+0.8z_{\mathrm{sepc}}\;\;\;(z_{\mathrm{sepc}} \leq 0.369), \\
      -0.169+1.8z_{\mathrm{sepc}}\;\;\;(0.369 < z_{\mathrm{sepc}} \leq 0.75), \\
      1.346-0.22z_{\mathrm{sepc}}\;\;\;(0.75 < z_{\mathrm{sepc}} \leq 1.1), \\
      0.004+z_{\mathrm{sepc}}\;\;\;(1.1 < z_{\mathrm{sepc}}), 
    \end{array}
  \right.
\label{eq:ri1}
\end{equation}
\begin{equation}
  r-i<
\left\{
    \begin{array}{l}
      0.5+1.35z_{\mathrm{sepc}}\;\;\;(z_{\mathrm{sepc}} \leq 0.8), \\
      1.98-0.5z_{\mathrm{sepc}}\;\;\;(0.8 < z_{\mathrm{sepc}} \leq 1.15), \\
      0.025+1.2z_{\mathrm{sepc}}\;\;\;(1.15 < z_{\mathrm{sepc}}), 
    \end{array}
  \right.
\end{equation}
\begin{equation}
  i-z>
\left\{
    \begin{array}{l}
      0.122+0.3z_{\mathrm{sepc}}\;\;\;(z_{\mathrm{sepc}} \leq 0.76), \\
      -1.132+1.95z_{\mathrm{sepc}}\;\;\;(0.76 < z_{\mathrm{sepc}} \leq 0.92), \\
      -0.074+0.8z_{\mathrm{sepc}}\;\;\;(0.92 < z_{\mathrm{sepc}} \leq 1.14), \\
       2.092-1.1z_{\mathrm{sepc}}\;\;\;(1.14 < z_{\mathrm{sepc}}),
    \end{array}
  \right.
\label{eq:iz1}
\end{equation}
\begin{equation}
  i-z<
\left\{
    \begin{array}{l}
      0.4+0.3z_{\mathrm{sepc}}\;\;\;(z_{\mathrm{sepc}} \leq 0.75), \\
      -0.8+1.9z_{\mathrm{sepc}}\;\;\;(0.75 < z_{\mathrm{sepc}} \leq 0.9), \\
      0.46+0.5z_{\mathrm{sepc}}\;\;\;(0.9 < z_{\mathrm{sepc}} \leq 1.17), \\
      2.098-0.9z_{\mathrm{sepc}}\;\;\;(1.17 < z_{\mathrm{sepc}}), 
    \end{array}
  \right.
\end{equation}
\begin{equation}
  z-y>
\left\{
    \begin{array}{l}
      0.02+0.1z_{\mathrm{sepc}}\;\;\;(z_{\mathrm{sepc}} \leq 0.9), \\
      -0.88+1.1z_{\mathrm{sepc}}\;\;\;(0.9 < z_{\mathrm{sepc}}), 
    \end{array}
  \right.
\end{equation}
\begin{equation}
  z-y<
\left\{
    \begin{array}{l}
      0.33+0.1z_{\mathrm{sepc}}\;\;\;(z_{\mathrm{sepc}} \leq 1.0), \\
      -0.67+1.1z_{\mathrm{sepc}}\;\;\;(1.0 < z_{\mathrm{sepc}}), 
    \end{array}
  \right.
\end{equation}
and use only spectroscopic galaxies with $i>17$ to avoid saturations
in the HSC-SSP images. We note that these color cuts are rough color
cuts before deriving the color-redshift relations with outlier
clippings and thus do not need to be strict \citep[see][]{oguri18a}.

\section{Analytic Model of Lensing Profiles}\label{app:halo_lensing}

While {\tt DarkEmulator} \citep{nishimichi19} has been used for the
previous analysis of the mass-richness relation in \citet{murata19},
here we take the so-called halo model approach to compute cluster
lensing profiles, largely following \citet{oguri11}. In this approach,
contributions from one-halo (1h) and two-halo (2h) terms are considered
separately. For the one-halo term, we adopt the Baltz-Marshall-Oguri
\citep[BMO;][]{baltz09,oguri11} radial density profile
\begin{equation}
\rho_{\mathrm{BMO}}(r)=\frac{\rho_{\mathrm{s}}}{(r/r_{\mathrm{s}})(1+r/r_{\mathrm{s}})^2}
\left(\frac{r_{\mathrm{t}}^2}{r^2+r_{\mathrm{t}}^2}\right)^n,
\label{eq:bmo}
\end{equation}
with $n=2$. This density profile is smoothly truncated at $r\simeq
r_{\mathrm{t}}$. The parameter $\rho_{\mathrm{s}}$ and $r_{\mathrm{s}}$ are
connected with the halo mass and the concentration parameter ignoring
the truncation \citep[see][]{oguri11}. We introduce the dimensionless
truncation radius defined by
$\tau_{200{\mathrm{m}}}=r_{\mathrm{t}}/r_{200{\mathrm{m}}}$, and
assume $\tau_{200{\mathrm{m}}}=2.5$ throughout the paper based on the
comparison with ray-tracing simulations presented in
\citet{oguri11}. For each halo mass and redshift, the concentration
parameter is calculated following \citet{diemer15} and \citet{diemer19}
using the {\tt colossus} package \citep{diemer18}. Explicit analytical
expressions of lensing profiles of the BMO profile,
$\Sigma_{\mathrm{BMO}}(R)$ and $\Delta\Sigma_{\mathrm{BMO}}(R)$, are
given in \citet{baltz09} and \citet{oguri11}.

We take account of the mis-centering effect in the same manner as in
\citet{murata19} that in fact follows a prescription proposed in
\citet{oguri11b}. More specifically, \citet{oguri11b} show that the
mis-centering PDF given by equation~(\ref{eq:poff}) with
$\sigma_1=0$ and $\sigma_2=\sigma_{\mathrm{off}}$ changes the Fourier
transform of the surface mass distribution $\tilde{\Sigma}_{\mathrm{BMO}}(k)$ as
\begin{equation}
  \tilde{\Sigma}^{\mathrm{off}}_{\mathrm{BMO}}(k)=
  \left[f_{\mathrm{cen}}+(1-f_{\mathrm{cen}})\exp\left(-\frac{k^2\sigma_{\mathrm{off}}^2}{2}\right)\right]\tilde{\Sigma}_{\mathrm{BMO}}(k).
\end{equation}
We note that an explicit analytic expression of the Fourier transform
of the normalized BMO profile, $\tilde{u}_{\mathrm{BMO}}(k)$, is given
in \citet{oguri11}, from which the Fourier transform of
$\Sigma_{\mathrm{BMO}}(R)$ is obtained as
$\tilde{\Sigma}_{\mathrm{BMO}}(k)=M_{200\mathrm{m}}\tilde{u}_{\mathrm{BMO}}(k)$.
From this expression in Fourier space, lensing profiles in real space
are obtained as
\begin{align}
  \Sigma^{\mathrm{off}}_{\mathrm{BMO}}(R) &=\int
  \frac{k\,dk}{2\pi}\tilde{\Sigma}^{\mathrm{off}}_{\mathrm{BMO}}(k)J_0(kR),
  \label{eq:ft_sigma}\\
 \Delta\Sigma^{\mathrm{off}}_{\mathrm{BMO}}(R) &=\int
 \frac{k\,dk}{2\pi}\tilde{\Sigma}^{\mathrm{off}}_{\mathrm{BMO}}(k)J_2(kR).
 \label{eq:ft_dsigma}
\end{align}
We use the {\tt FFTLog} algorithm \citep{hamilton00} implemented by
\citet{fang20} and also used in {\tt DarkEmulator}
\citep{nishimichi19} for fast calculations of the inverse Fourier
transform. 

\begin{figure}
 \begin{center}
   \includegraphics[width=8.6cm]{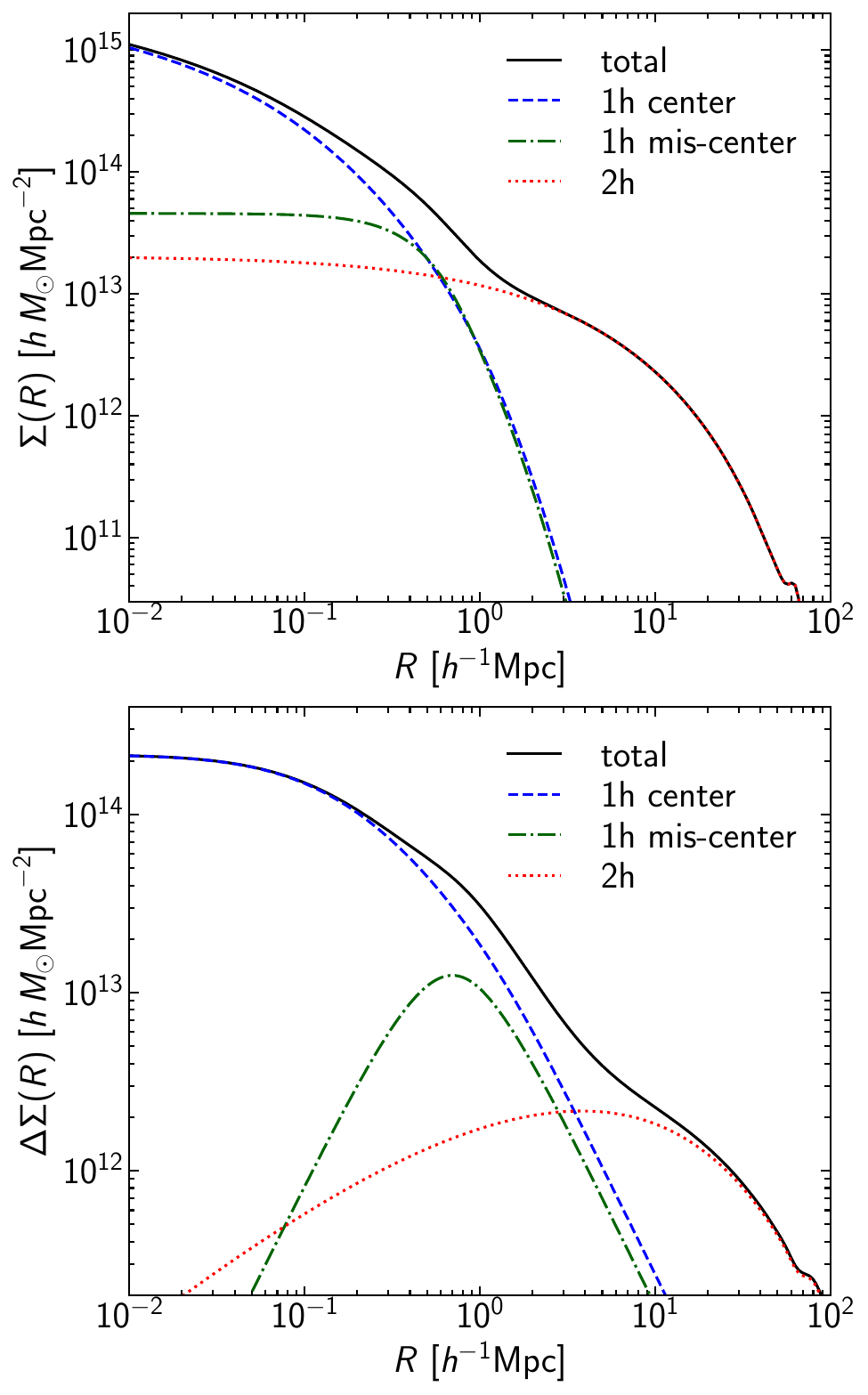} 
 \end{center}
 \caption{Example of lensing profiles $\Sigma(R)$ ({\it upper}) calculated by
   equation~(\ref{eq:halo_sigma_tot}) and $\Delta\Sigma(r)$ ({\it
     lower}) calculated by equation~(\ref{eq:halo_dsigma_tot}).
   We assume the halo mass $M_{200{\mathrm{m}}}=1\times 10^{14}h^{-1}M_\odot$, the
  redshift $z=0.5$, the mis-centering parameters
  $f_{\mathrm{cen}}=0.6$ and $\sigma_{\mathrm{off}}=0.3h^{-1}$Mpc, and
  the cosmological parameters shown in Sec.~\ref{sec:stacked_lensing}.
  Note that we show lensing profiles in the physical length
  scale. Dashed, dash-dotted, and dotted lines show contributions of
  each term described in the text to the total lensing profiles.
    {Alt text: In each panel, the x axis shows the distance from the
      cluster center from 0.01 to 100 megaparsec. The y axis shows the
      differential surface mass distribution.}
}\label{fig:dsigma_example}  
\end{figure}

Following \citet{oguri11} and \citet{oguri11b}, the Fourier transform
of the two-halo term is computed as
\begin{equation}
  \tilde{\Sigma}^{\mathrm{2h}}(k)
  =\bar{\rho}_{\mathrm{m0}}b_{\mathrm{h}}(M)P_{\mathrm{m}}(k/(1+z);\,z),
\end{equation}
where $\bar{\rho}_{\mathrm{m0}}$ is the average matter density of the
Universe at $z=0$, $b_{\mathrm{h}}(M)$ is the halo bias
\citep{tinker10}, and $P_{\mathrm{m}}(k/(1+z);\,z)$ is the linear
matter power spectrum. The division of the wavenumber by $(1+z)$ in
the linear matter power spectrum originates from the fact the
wavenumber in the power spectrum is usually defined in the comoving
scale, while $k$ in this equation is in the physical scale. 
The corresponding two-halo lensing profiles in real space,
$\Sigma^{\mathrm{2h}}(R)$ and $\Delta\Sigma^{\mathrm{2h}}(R)$, are derived
in the same manner as in equations~(\ref{eq:ft_sigma}) and
(\ref{eq:ft_dsigma}). Total lensing profiles are obtained by summing
up one-halo and two-halo terms as
\begin{align}
  \Sigma(R)&=\Sigma^{\mathrm{off}}_{\mathrm{BMO}}(R)+\Sigma^{\mathrm{2h}}(R),
  \label{eq:halo_sigma_tot}\\
  \Delta\Sigma(R)&=\Delta\Sigma^{\mathrm{off}}_{\mathrm{BMO}}(R)+\Delta\Sigma^{\mathrm{2h}}(R).
  \label{eq:halo_dsigma_tot}
\end{align}
The {\tt python} code to compute these lensing profiles is made
available at {\tt https://github.com/oguri/halo\_lensing}. An example of lensing profiles computed with this
model is presented in Figure~\ref{fig:dsigma_example}.

 \bibliographystyle{apj}
 \bibliography{refs}

\end{document}